%% file: main.tex
\documentclass[a4paper,english,cleveref,autoref]{lipics-v2019}

\input{mymacros}

\usepackage{microtype}
\title{Boolean Tensor Decomposition for Conjunctive Queries with Negation}

\titlerunning{Boolean Tensor Decomposition for Conjunctive Queries with Negation}

\author{Mahmoud Abo Khamis}{RelationalAI, Berkeley, USA}{}{}{}

\author{Hung Q. Ngo}{RelationalAI, Berkeley, USA}{}{}{}

\author{Dan Olteanu}{Department of Computer Science, University of Oxford, Oxford, UK}{}{}{}

\author{Dan Suciu}{Department of Computer Science and Engineering, University of Washington, USA}{}{}{}

\authorrunning{M. Abo Khamis, H.Q. Ngo, D. Olteanu, and D. Suciu}

\Copyright{Mahmoud Abo Khamis, Hung Q. Ngo, Dan Olteanu, and Dan Suciu}



\funding{This project has received funding from the European Union's Horizon 2020 research and innovation programme under grant agreement No 682588.
This project is also supported in part by NSF grants AITF-1535565 and III-1614738.}

\acknowledgements{The authors would like to thank the anonymous reviewers for their suggestions that helped improve the readability of this paper.}

\nolinenumbers 

\hideLIPIcs

\EventEditors{Pablo Barcelo and Marco Calautti}
\EventNoEds{2}
\EventLongTitle{22nd International Conference on Database Theory (ICDT 2019)}
\EventShortTitle{ICDT 2019}
\EventAcronym{ICDT}
\EventYear{2019}
\EventDate{March 26--28, 2019}
\EventLocation{Lisbon, Portugal}
\EventLogo{}
\SeriesVolume{127}
\ArticleNo{18}

\begin{document}

\maketitle

\begin{abstract}
\input{abstract}

\end{abstract}

\input{intro}

\input{prelim}

\input{example}

\input{results}
\input{relatedwork}

\input{conclusions}


\bibliography{sample}

\appendix
\input{appendix}

\end{document}

%% file: mymacros.tex
\usepackage{bm}
\usepackage[dvipsnames]{xcolor}
\usepackage{cite,color,xspace}
\usepackage{algorithm}
\usepackage[noend]{algpseudocode}

\usepackage{tikz,tikz-qtree,tikz-qtree-compat}
\usetikzlibrary{shapes,snakes,arrows,calc}

\newcommand{\yell}[1]{{\color{red} \textbf{#1}}}

\newcommand{\pr}{\mathop{\textnormal{Prob}}}    
\newcommand{\dom}{\textsf{Dom}}
\newcommand{\skeleton}{\textsf{body}}
\newcommand{\fp}{\textsf{FP}}

\newcommand{\nae}{\textnormal{\sf NAE}}

\newcommand{\tw}{\textnormal{\sf tw}}
\newcommand{\itw}{\textnormal{\sf itw}}
\newcommand{\ifhtw}{\textnormal{\sf ifhtw}}

\newcommand{\SWONE}{\#\textnormal{\sf W}[1]}

\newcommand{\subw}{\textnormal{\sf subw}}
\newcommand{\panda}{\textnormal{\sf PANDA}}

\newcommand{\Dom}{\functionname{Dom}}
\newcommand{\D}{\mathbf D}

\newcommand{\outputsize}{|\textnormal{\sf output}|}

\newcommand{\agm}{\text{\sf AGM}}
\newcommand{\bi}{\begin{itemize}}
\newcommand{\ei}{\end{itemize}}
\newcommand{\bp}{\begin{proof}}
\newcommand{\ep}{\end{proof}}
\newcommand{\bcor}{\begin{cor}}
\newcommand{\ecor}{\end{cor}}
\newcommand{\bthm}{\begin{thm}}
\newcommand{\ethm}{\end{thm}}
\newcommand{\blmm}{\begin{lmm}}
\newcommand{\elmm}{\end{lmm}}
\newcommand{\bdefn}{\begin{defn}}
\newcommand{\edefn}{\end{defn}}
\newcommand{\bprop}{\begin{prop}}
\newcommand{\eprop}{\end{prop}}
\newcommand{\bconj}{\begin{conj}}
\newcommand{\econj}{\end{conj}}
\newcommand{\bopm}{\begin{opm}}
\newcommand{\eopm}{\end{opm}}
\newcommand{\brmk}{\begin{rmk}}
\newcommand{\ermk}{\end{rmk}}

\newcommand{\functionname}[1]{\text{\sf #1}}
\newcommand{\faq}{\text{\sf FAQ}}
\newcommand{\sumprod}{\text{\sf SumProd}}
\newcommand{\InsideOut}{\functionname{InsideOut}}
\newcommand{\true}{\textsf{true}}
\newcommand{\false}{\textsf{false}}

\newcommand{\nop}[1]{}

\newcommand{\R}{\mathbb R}
\newcommand{\RS}{\text{\sf RS}}
\newcommand{\F}{\mathbb F}
\newcommand{\poly}{\text{\sf poly}}
\newcommand{\norm}[1]{\|#1\|}
\newcommand{\calH}{\mathcal H}
\newcommand{\calF}{\mathcal F}
\newcommand{\calS}{\mathcal S}
\newcommand{\calT}{\mathcal T}

\newcommand{\col}{\text{\sf col}}
\newcommand{\calA}{\mathcal A}
\newcommand{\calG}{\mathcal G}
\newcommand{\calD}{\mathcal D}
\newcommand{\calP}{\mathcal P}

\newcommand{\calV}{\mathcal V}
\newcommand{\calE}{\mathcal E}
\newcommand{\fhtw}{\text{\sf fhtw}}
\newcommand{\np}{\textnormal{\sf NP}}
\newcommand{\ptime}{\textnormal{\sf PTIME}}
\newcommand{\suchthat}{\ | \ }

\theoremstyle{definition}              
\newtheorem{thm}{Theorem}[section]
\newtheorem{lmm}[thm]{Lemma}
\newtheorem{prop}[thm]{Proposition}
\newtheorem{cor}[thm]{Corollary}

\newtheorem{pbm}{Problem}
\newtheorem{opm}{Question}
\newtheorem{conj}[thm]{Conjecture}
\newtheorem{ex}[thm]{Example}

\newtheorem{defn}[thm]{Definition}

\newtheorem{rmk}[thm]{Remark}

\newcommand{\ed}{\text{\sf ED}}

\newcommand{\td}{\textsf{TD}}

%% file: abstract.tex
We propose an approach for answering conjunctive queries with negation, where
the negated relations have bounded degree. Its data complexity matches that of the $\InsideOut$ and $\panda$ algorithms for the positive subquery of the input query and is expressed in terms of the fractional hypertree width and the submodular width respectively. Its query complexity depends on the structure of the conjunction of negated relations; in general it is exponential in the number of join variables occurring in negated relations yet it becomes polynomial for several classes of queries.

This approach relies on several contributions. We show how to rewrite queries with negation on bounded-degree relations into equivalent conjunctive queries with not-all-equal ($\nae$) predicates, which are a multi-dimensional analog of disequality ($\neq$). We then generalize the known color-coding technique to conjunctions of $\nae$ predicates and explain it via a Boolean tensor decomposition of conjunctions of $\nae$ predicates. This decomposition can be achieved via a probabilistic construction that can be derandomized efficiently.

%% file: intro.tex
\section{Introduction}
\label{sec:intro}

This paper considers the problem of answering conjunctive queries with negation of the form
\begin{equation}
   Q(\bm X_F) \leftarrow 
   \skeleton \wedge \bigwedge_{S\in\overline\calE} \neg R_S(\bm X_S),
   \label{eqn:Q:R:F}
\end{equation}
where $\skeleton$ is the body of an arbitrary conjunctive query, 
$\bm X_F = (X_i)_{i\in F}$ denotes a tuple
of variables (or attributes) indexed by a set $F$ of positive integers,
and $\overline\calE$ is the set of
hyperedges of a multi-hypergraph\footnote{In a multi-hypergraph, each 
hyperedge $S$ can occur multiple times. All hypergraphs in this paper are multi-hypergraphs.}
$\overline\calH = (\overline\calV,\overline\calE)$.
Every hyperedge $S \in \overline\calE$ corresponds to a bounded-degree relation $R_S$ on  attributes $\bm X_S$. 
For instance, the equality ($=$) relation is a
bounded-degree (binary) relation, because every element in the active domain
has degree one; the edge relation $E$ of a graph with bounded maximum
degree is also a bounded-degree relation.
Section~\ref{sec:prelim} formalizes this notion of bounded degree. 

\nop{
Eq.~\eqref{eqn:Q:R:F} captures SQL queries with NOT EXISTS clau\-ses and disequality predicates ($\neq$). In the rule mining problem~\cite{DBLP:conf/sigmod/ChenGWJ16}, 
one must count the number of violations of a conjunctive rule, 
which also leads to Eq.~\eqref{eqn:Q:R:F}.}
We exemplify  using three Boolean queries\footnote{We denote Boolean queries $Q(\bm X_F)$ where $F=\emptyset$ by $Q$ instead of $Q()$. We also use $[n] =\{1,\dots,n\}$.
} over a directed graph $G = ([n],E)$ with $n$ nodes and $N=|E|$ edges:
the $k$-walk query\footnote{Unlike a path, in a walk  some vertices may
repeat.}, the $k$-path query, and the induced (or chordless) $k$-path query.
They have the same $\skeleton$ and encode graph problems of increasing
complexity: 
{\small
\begin{align}
   W \leftarrow &E(X_1,X_2) \wedge E(X_2,X_3) \wedge \cdots \wedge E(X_{k},X_{k+1}).\nonumber\\
   P \leftarrow &E(X_1,X_2) \wedge E(X_2,X_3) \wedge \dots \wedge E(X_{k},X_{k+1})      
        \wedge \bigwedge_{\substack{i,j\in [k+1]\\ i+1<j}}X_i \neq X_j.
        \label{eq:intro:Q2}\\
   I \leftarrow &E(X_1,X_2) \wedge E(X_2,X_3) \wedge \dots \wedge E(X_{k},X_{k+1}) 
       \wedge \bigwedge_{\substack{i,j\in [k+1] \\ i+1<j}} (\neg E(X_i, X_j) \wedge
         X_i \neq X_j)). \label{eq:intro:Q3}
\end{align}
}
The hypergraph $\overline\calH$ for the $k$-walk query $W$ is empty since it has no negated relations. This query
can be answered in $O(kN\log N)$ time using for instance the Yannakakis dynamic programming algorithm~\cite{Yannakakis:VLDB:81}. 
The $k$-path query $P$ has the hypergraph $\overline \calH = ([k+1],
\{ (i,j) \suchthat i,j\in [k+1], i+1<j\})$.
It can be answered in $O(k^kN\log N)$-time~\cite{PlehnV90} and 
even better in $2^{O(k)}N\log N$-time using the color-coding
technique~\cite{DBLP:journals/jacm/AlonYZ95}.
The induced $k$-path query $I$ has the hypergraph $\overline \calH$ similar to that of $P$, but every edge $(i,j)$ has now multiplicity two due to the negated edge relation and also the disequality.
This query is $\mathsf{W}[2]$-hard~\cite{DBLP:conf/coco/ChenF07}.
However, if the graph $G$ has a maximal degree that is bounded by some constant $d$, then the query can be answered in 
$O(f(k, d)\cdot N \log N)$-time for some function $f$ that depends exponentially on $k$ and $d$~\cite{PlehnV90}.
Our results imply the above complexities for the three queries.

\subsection{Main Contribution}


In this paper we propose an approach to answering conjunctive queries with negation on bounded-degree relations of arbitrary arities. 
Our approach is the first to exploit the bounded degree of the negated relations. The best known algorithms for positive queries such as $\InsideOut$~\cite{faq} and $\panda$~\cite{panda} can also answer queries with negation, albeit with much higher complexity since already one negation can increase their worst-case runtime. For example, the Boolean path queries with a disequality between the two end points takes linear time with our approach, but quadratic time with existing approaches~\cite{faq,panda}.
The data complexity of our approach matches that of $\InsideOut$ and $\panda$ for the positive subquery $Q(\bm X_F) \leftarrow \skeleton$. To lower its query complexity, we use a range of techniques including color-coding, probabilistic construction of Boolean tensor decompositions, and derandomization of this construction.

\bthm
Any query $Q$ of the form~\eqref{eqn:Q:R:F}, where for each $S\in\overline\calE$ the relation $R_S$ has bounded degree and $f(\cdot)$ is a function of $Q$, can be answered over a database of size $N$ 
in time $O(f(Q)\cdot\log N\cdot(N^{\fhtw_F(\skeleton)}+\outputsize))$ using a reduction to $\InsideOut$ and $O(f(Q)\cdot(\poly(\log N)\cdot N^{\subw_F(\skeleton)}+\log N\cdot \outputsize))$ using a reduction to $\panda$. 
\label{thm:complexity}
\ethm

The complexities of $\InsideOut$~\cite{faq} and $\panda$~\cite{panda} depend on the 
fractional hypertree width $\fhtw$~\cite{DBLP:journals/talg/GroheM14} and respectively the submodular width $\subw$~\cite{Marx:subw}.
The widths $\fhtw_F$ and $\subw_F$ are $\fhtw$ and respectively $\subw$ computed on the subset of hypertree decompositions of the positive subquery of $Q$ for which the set $F$ of free variables form a connected subtree. The dependency of the function $f$ on the structure of $Q$, and in particular on the hypergraph  $\overline\calH$ of the negated relations in $Q$, is an important result of this paper.

\nop{
Its query complexity is at most that of known special cases and depends on the query structure. 

In general, the query complexity depends exponentially on the number $p$ of negated relations.
Our algorithm is sensitive to the query structure and benefits from two natural ideas to reduce the query complexity. Due to space limitation, we defer to Appendix~\ref{sec:query:complexity} a discussion on how to lower the query complexity by exploiting the query structure.
}

Theorem~\ref{thm:complexity} draws on three contributions: 
\begin{enumerate}
\item A rewriting of queries of the form~\eqref{eqn:Q:R:F} into equivalent conjunctive queries with not-all-equal predicates, which are a multi-dimensio\-nal analog of disequality $\neq$ (Proposition~\ref{prop:reduced:form}); 

\item A generalization of color-coding~\cite{DBLP:journals/jacm/AlonYZ95} from cliques of disequalities to arbitrary conjunctions of not-all-equal predicates; and 

\item An alternative view of color coding via Boolean tensor decomposition of conjunctions of not-all-equal predicates (Lemma~\ref{lemma:boolean-tensor-decomposition}). This decomposition admits a probabilistic construction that can be derandomized efficiently (Corollary~\ref{cor:bounding-boolean-rank}).

\end{enumerate}

Contribution 1 (Section~\ref{sec:results}) gives a rewrite of the query $Q$ into an equivalent disjunction of queries
$Q_i$ of the form (cf.\@ Proposition~\ref{prop:reduced:form})
\[ Q_i(\bm X_F) \leftarrow \skeleton_i \wedge \bigwedge_{S \in
\calE_i} \nae(\bm Z_S). 
\]

For each query $Q_i$, $\skeleton_i$ may be different from
$\skeleton$ in $Q$, since fresh variables $\bm Z_S$ and unary predicates may be introduced. Its fractional hypertree and submodular widths remain however at most that of $\skeleton$. 
We thus rewrite the conjunction of the negated relations into a much simpler conjunction of $\nae$ predicates without increasing the data complexity of $Q$. The number of such queries $Q_i$ depends exponentially on the arities and the degrees of the negated relations, which is the reason why we need the constant bound on these degrees.

Contribution 2 (Section~\ref{subsec:boolean:tensor:decomp}) is based on the observation that a conjunction of $\nae$ predicates can be answered by 
an adaptation of the color-coding technique~\cite{DBLP:journals/jacm/AlonYZ95}, which has been used so far for checking cliques of disequalities. The crux of this technique is to randomly color each value in the active domain with one color from a set whose size is much smaller than the size of the active domain, and to use these colors instead of the values themselves to check the disequalities. We generalize this idea to conjunctions of $\nae$ predicates and show that such conjunctions can be expressed equivalently as disjunctions of simple queries over the different possible colorings of the variables in these queries.

Contribution 3 (Section~\ref{subsec:boolean:tensor:decomp}) explains color coding by providing an alternative view of it: 
Color coding is a (Boolean) tensor decomposition of
the (Boolean) tensor defined by the conjunction $\bigwedge_S \nae(\bm Z_S)$. 
As a tensor, $\bigwedge_S \nae(\bm Z_S)$ is a multivariate function over
variables in the set $U = \bigcup_S\bm Z_S$. 
The tensor decomposition rewrites it into a disjunction of conjunctions
of {\em univariate} functions over individual variables $Z_i$ (Lemma~\ref{lmm:decomp:id}). That is,
\[\bigwedge _S \nae(\bm Z_S) \equiv \bigvee_{j\in [r]}\bigwedge_{i\in U} f_{i}^{(j)}(Z_i),\]

where $r$ is the (Boolean tensor) rank of the tensor decomposition, and for each $j\in[r]$, the inner conjunction $\bigwedge_{i\in U} f_{i}^{(j)}(Z_i)$ can be thought of as a rank-1 tensor of inexpensive Boolean univariate functions $f_{i}^{(j)}(\cdot)$ ($\forall i\in U$).
The key advantages of this tensor decomposition are that (i) the addition of univariate conjuncts to $\skeleton_i$ does not increase its (fractional hypertree and submodular) width and (ii) the dependency of the rank $r$ on the database size $N$ is only a $\log N$ factor. Lemma~\ref{lmm:decomp:id} shows that the rank $r$ depends on two quantities: $r = P(\calG, c) \cdot |\calF|$. The first is the chromatic polynomial of the hypergraph of $\bigwedge_S \nae(\bm Z_S)$ using $c$ colors. The second is the size of a family of hash functions that represent proper $c$-colorings of homomorphic images of the input database. The number $c$ of needed colors is at most the number $|U|$ of variables in $\bigwedge_S \nae(\bm Z_S)$. We show it to be the maximum chromatic number of a hypergraph defined by any homomorphic image of the database.

We give a probabilistic construction of the tensor decomposition that generalizes the construction used by the color-coding technique. 
It selects a color distribution dependent on the query structure, which allows the rank of $\bigwedge_S \nae(\bm Z_S)$ to take a wide range of query complexity asymptotics, from polynomial to exponential in the query size.
This is more refined than the previously known bound~\cite{DBLP:journals/jacm/AlonYZ95}, which amounts to a tensor rank that is exponential in the query size. 
We further derandomize this construction by adapting ideas from derandomization for $k$-{\sf restrictions}~\cite{DBLP:journals/talg/AlonMS06} (with $k$ being related to the Boolean tensor rank).

Section~\ref{sec:use-tensor-decomposition} shows how to use the Boolean tensor decomposition in conjunction with $\InsideOut$~\cite{faq} and $\panda$~\cite{panda} to evaluate queries of the form~\eqref{eqn:Q:R:F} with the complexity given by Theorem~\ref{thm:complexity}. The query complexity captured by the function $f$ is given by the number of $\nae$ predicates and the rank of the tensor decomposition of their conjunction. 

\nop{Our approach further shaves off a $\log N$ factor in the number of colors used for color coding.
Recall that the RAM model of computation comes in two variants~\cite{DBLP:books/cu/MotwaniR95}: the bit and the unit models, where the cost  of a single operation is defined to be $\log N$  and $1$ respectively.  We show that the operations on $\log N$ colors can be encoded as bit operations that only take $1$ step in the unit model.}

%% file: prelim.tex
\section{Preliminaries}
\label{sec:prelim}

In this paper we consider arbitrary conjunctive queries with negated relations of the form \eqref{eqn:Q:R:F}. 
We make use of the following naming convention. Capital letters
with subscripts such as $X_i$ or $A_j$ denote variables. For any set $S$ of positive
integers, $\bm X_S = (X_i)_{i\in S}$ denote a tuple of variables indexed by $S$.
Given a relation $R$ over variables $\bm X_S$ and $J\subseteq S$,
$\pi_J R$ denotes the projection of $R$ onto variables $\bm X_J$, i.e.,
we write $\pi_J R$ instead of $\pi_{\bm X_J} R$.
If $X_i$ is a variable, then the corresponding lower-case $x_i$ denotes a value
from the active domain $\dom(X_i)$ of $X_i$.
Bold-face $\bm x_S = (x_i)_{i\in S}$ denotes a tuple of values in 
$\prod_{i\in S} \dom(X_i)$.

We associate a hypergraph $\calH(R)$ with a relation $R(\bm X_S)$ as follows. 
The vertex set is $\{ (i, v) \ | \ i \in S, v \in \dom(X_i)\}$. 
Each tuple $\bm x_S = (x_i)_{i\in S} \in
R$ corresponds to a (hyper)edge $\{ (i,x_i) | i \in S\}$.
$\calH(R)$ is a $|S|$-uniform hypergraph (all hyperedges have size $|S|$).

\subsection{Hypergraph coloring and bounded-degree relations}
\subparagraph{Hypergraph coloring.}
Let $\calG = (U,\calA)$ denote a multi-hypergraph 
and $k$ be a positive integer.
A {\em proper  $c$-coloring} of $\calG$ is a mapping $h : U \to [c]$ such that
for every edge $S \in \calA$, there exists $u, v\in S$ with $u\neq v$
such that $h(u)\neq h(v)$.
The {\em chromatic polynomial} $P(\calG, c)$
of  $\calG$ is the number of proper $c$-colorings of
$\calG$~\cite{MR95h:05067}.
A vertex (edge) coloring of $\calG$ is an assignment of colors to the vertices  (edges) of $\calG$ so that no two adjacent vertices (incident edges) have the same color. The {\em chromatic number} $\chi(\calG)$ and the {\em chromatic index} $\chi'(\calG)$ are the smallest numbers of colors needed for a vertex coloring and respectively an edge coloring of $\calG$.
Coloring a (hyper)graph is equivalent to coloring it without singleton edges.

\subparagraph{Bounded-degree relation.}
The {\em maximum degree} of a vertex in a hypergraph $\calG = (U,\calA)$ is denoted by $\Delta(\calG)$:
 $\Delta(\calG)= \max_{v\in U}|\{S\in\calA \suchthat v \in S\}|$. For a relation $R_S(\bm X_S)$, its 
maximum degree $\Delta(\calH(R_S))$ is the maximum number of tuples in $R_S$ with the same value for a variable $X \in \bm X_S$: 
$\Delta(\calH(R_S))=\max_{\substack{i\in S\\v\in\dom(X_i)}}|\{\bm x_S\in R_S \suchthat x_i=v\}|.$
We will use a slightly different notion of {\em degree} of a relation denoted by $\deg(R_S)$, which also accounts for the arity $|S|$ of the relation $R_S$. Proposition~\ref{prop:matching:decomp} connects the two notions.

\bdefn[Matching]
A $k$-ary relation $M(\bm X_S)$ is called a ($k$-dimensional) {\em matching}
if for every two tuples $\bm x_S, \bm x'_S \in M$, either $\bm x_S=\bm x'_S$, i.e., $\bm x_S$ and $\bm x'_S$ are the same tuple, or it holds that
$x_i \neq x'_i, \forall i \in S$.\label{def:matching}
\edefn

\bdefn[Degree]
The {\em degree} of a relation $R_S(\bm X_S)$, denoted by $\deg(R_S)$,
is the smallest integer $d$ for which $R_S$ can be written as the disjoint union 
of $d$ matchings. The degree $\deg(R_S)$ is {\em bounded} if there is a constant $d_S$ such that $\deg(R_S)\leq d_S$.
\label{defn:degR}
\edefn

It is easy to see that $\deg(R_S) = \chi'(\calH(R_S))$.
If $R_S$ is a binary relation, then $\calH(R_S)$ is a bipartite graph
and $\deg(R_S) = \chi'(\calH(R_S))=\Delta(\calH(R_S))$.
This follows from K\"onig's line coloring theorem~\cite{konig1916},
which states that the chromatic index of a bipartite graph is equal to its
maximum degree. 
When the arity $k$ is higher than two, to the best of our 
knowledge there does not exist such a nice characterization of the chromatic
index of $R_S$ in terms of the maximum degree
of individual vertices in the graph, although there has been some work on
bounding the chromatic index of (linear) uniform hypergraphs~\cite{MR1426745,MR3324967,
2016arXiv160304938F,MR993646, MR993646}. 
In our setting, we are willing to live with sub-optimal decomposition of a bounded-degree relation into matchings as long as it can be done in linear time.

\bprop
Let $R_S(\bm X_S)$ denote a $k$-ary relation of size $N$ and $\ell=\Delta(\calH(R_S))$. Then:
\bi
 \item $\ell \leq \deg(R_S) \leq k(\ell-1)+1$;
 \item We can compute in $O(N)$-time disjoint $k$-ary matchings $M_1, \dots,
    M_{k\ell-k+1}$ such that $R_S = \bigcup_{j=1}^{k(\ell-1)+1} M_j$.
\ei
\label{prop:matching:decomp}
\eprop
\begin{proof}
The fact that $\ell \leq \deg(R_S)$ is obvious. To show that $\deg(R_S)\leq k(\ell-1)+1$, note
that any edge in $\calH(R_S)$ is adjacent to at most $k(\ell-1)$ other edges of $\calH(R_S)$, hence greedy coloring can color the edges of $\calH(R_S)$
in time $O(N)$ using $k(\ell-1)+1$ colors.
\end{proof}

The two notions of degree of a relation are thus equivalent up to a constant factor given by the arity of the relation.

\subsection{$\faq$, width parameters, and corresponding algorithms}
\label{subsec:insideout-panda}
\bdefn[The $\faq$ problem~\cite{faq}]
The input to $\faq$ is a set of functions and the output is a function which is a series of aggregations (e.g. sums) over the product of input functions.
In particular, the input to $\faq$ consists of the following:
\bi
\item A multi-hypergraph $\calH=(\calV=[n], \calE)$.
\item Each vertex $i \in \calV = [n]$ corresponds to a variable $X_i$ over a discrete domain $\Dom(X_i)$.
\item For each hyperedge $S\in\calE$, there is a corresponding input function (also called a {\em factor}) $\psi_S:\prod_{i\in S} \Dom(X_i) \to \D$ for some fixed $\D$.
\item A number $f\in[n]$. Let $F:=[f]$. Variables in $\bm X_F$ are called {\em free variables}, while variables in $\bm X_{[n]-F}$ are called {\em bound variables}.
\item For each $i\in[n]-F$, there is a commutative semiring $(\D, \bigoplus^{(i)}, \bigotimes)$. (All the semirings share the same $\D$ and $\bigotimes$ but can potentially have different $\bigoplus^{(i)}$).\footnote{More generally, instead of $(\D, \oplus^{(i)}, \otimes)$ being a semiring, we also allow some $\oplus^{^{(i)}}$ to be identical to $\otimes$.}
\ei
The $\faq$ problem is to compute the following function $\varphi(\bm x_{F}):\prod_{i\in F} \Dom(X_i) \to \D$
\begin{equation}
\varphi(\bm x_{F}) := \bigoplus_{x_{f+1}}^{(f+1)}\ldots\bigoplus_{x_{n}}^{(n)} \bigotimes_{S\in\calE} \psi_S(\bm x_s).
\end{equation}
\qed
\edefn

Consider the conjunctive query 
$Q(\bm X_F) \leftarrow \bigwedge_{S \in \calE} R_S(\bm X_S)$ where $\bm X_F$ is the set of free variables.
The $\faq$ framework models each input relation $R_S$ as a Boolean function
$\psi_S(\bm x_S)$, called a ``factor'', in which $\psi_S(\bm x_S)=\true$ iff $\bm x_S \in R_S$.
Then, computing the output $Q(\bm X_F)$ is equivalent to computing the Boolean function $\varphi(\bm x_F)$ defined as
$\varphi(\bm x_F) = \bigvee_{x_{f+1}} \cdots \bigvee_{x_n} \bigwedge_{S\in\calE}\psi_S(\bm
x_S).$
Instead of Boolean functions, this expression can be defined 
in $\sumprod$ form over
functions on a commutative semiring $(\bm D,\oplus,\otimes)$: 
\begin{equation}
\varphi(\bm x_F) = \bigoplus_{x_{f+1}} \cdots \bigoplus_{x_n} \bigotimes_{S\in\calE}\psi_S(\bm x_S). 
\label{eqn:sum:prod}
\end{equation}
The semiring $(\{\true,\false\},\vee,\wedge)$ was used for $Q$ above.

We next define tree decompositions and the $\fhtw$ and $\subw$ parameters. We refer the reader to the recent survey by
Gottlob et al.~\cite{DBLP:conf/pods/GottlobGLS16} for more details and a historical
context.
In what follows, the hypergraph $\calH$ should be thought of as the hypergraph of
the input $\faq$
query, although the notions of tree decomposition and width parameters are defined
independently of queries.

\bdefn[Tree decomposition]A {\em tree decomposition} of a hypergraph $\calH=(\calV,\calE)$
is a pair
$(T,\chi)$, where $T$ is a tree and $\chi: V(T) \to 2^{\calV}$ maps each node
$t$ of the tree to a subset $\chi(t)$ of vertices such that:
\bi
    \item[(a)] Every hyperedge $S\in \calE$ is a subset of some $\chi(t)$, $t\in V(T)$
    (i.e. every edge is covered by some bag);
    \item[(b)] For every vertex $v \in \calV$,
    the set $\{t \suchthat v \in \chi(t)\}$ is a non-empty (connected) sub-tree of $T$.
    This is called the {\em running intersection property}.
\ei
The sets $\chi(t)$ are often called the {\em bags}
of the tree decomposition.
Let $\td(\calH)$ denote the set of all tree decompositions of $\calH$.
When $\calH$ is clear from context, we use $\td$ for brevity.
\label{defn:td}
\edefn

\bdefn[$F$-connex tree
decomposition~\cite{Bagan:CSL:07,Segoufin:2013:ECD:2448496.2448498}]
\label{defn:F-connex}
Given a hypergraph $\calH=(\calV,\calE)$ and a set $F\subseteq \calV$,
a tree decomposition $(T,\chi)$ of $\calH$ is {\em $F$-connex} if there is a  subset $V'\subseteq V(T)$ that
forms a connected subtree of $T$ and satisfies $\bigcup_{t\in V'}\chi(t)=F$.
We use $\td_F$ to denote the set of all $F$-connex tree decompositions
of $\calH$. (Note that when $F=\emptyset$, $\td_F=\td$.)
\edefn

To define width parameters, we use the polymatroid characterization
from \cite{panda}.
A function $f : 2^{\calV} \to \R_+$ is called a (non-negative)
{\em set function} on $\calV$.
A set function $f$ on $\calV$ is {\em modular} if
$f(S) = \sum_{v\in S} f(\{v\})$ for all $S\subseteq \calV$,
it is {\em monotone} if $f(X) \leq f(Y)$ whenever $X \subseteq Y \subseteq \calV$,
and it is {\em submodular} if $f(X\cup Y)+f(X\cap Y)\leq f(X)+f(Y)$
for all $X,Y\subseteq \calV$.
A monotone, submodular set function $h : 2^{\calV} \to \R_+$ with $h(\emptyset)
= 0$ is called a {\em polymatroid}.
Let $\Gamma_n$ denote the set of all polymatroids on $\calV=[n]$.

Given a hypergraph $\calH=(\calV, \calE)$, define the set of {\em edge dominated} set
functions, denoted by $\ed_\calH$ or $\ed$ when $\calH$ is clear from the context, as follows:
\begin{align}
    \ed &:= \{ h \suchthat h : 2^{\calV} \to \R_+, h(S) \leq 1,
    \forall S \in \calE\}.\label{eqn:ed}
\end{align}
We can now define the submodular width and fractional hypertree width of
a given hypergraph $\calH$ (or of a given $\faq$ query with hypergraph $\calH$):
\footnote{Although this definition for $\fhtw$ differs from the original one~\cite{DBLP:journals/talg/GroheM14, DBLP:conf/pods/GottlobGLS16}, the two definitions have been shown to be equivalent~\cite{panda}.}
\begin{eqnarray*}
    \fhtw(\calH) := \min_{(T,\chi) \in \td} \max_{h \in \ed \cap \Gamma_n} 
    \max_{t\in V(T)}h(\chi(t)),& 
    \displaystyle{\fhtw_F(\calH) := \min_{(T,\chi) \in \td_F} \max_{h \in \ed \cap \Gamma_n} 
    \max_{t\in V(T)}h(\chi(t))},\\
    \subw(\calH) :=  \max_{h \in \ed \cap \Gamma_n}\min_{(T,\chi) \in \td} 
    \max_{t\in V(T)}h(\chi(t)),& 
    \displaystyle{\subw_F(\calH) :=  \max_{h \in \ed \cap \Gamma_n}\min_{(T,\chi) \in \td_F} 
    \max_{t\in V(T)}h(\chi(t)).}
\end{eqnarray*}
It is known that $\subw(\calH) \leq \fhtw(\calH)$, and there are
classes of
hypergraphs with bounded $\subw$ and unbounded $\fhtw$~\cite{Marx:subw}.
Furthermore, $\fhtw$ is
strictly less than other width notions such as (generalized) hypertree width and
tree width.

\bthm[\cite{faq}]
Given an $\faq$ $\varphi$ over a single semiring with hypergraph $\calH=(\calV, \calE)$ and free variables $F\subseteq \calV$ over a database of size $N$, the $\InsideOut$ algorithm can answer $\varphi$ in time $O(|\calE|\cdot|\calV|^2\cdot \log N\cdot (N^{\fhtw_F(\calH)} + \outputsize))$.
\label{thm:insideout-runtime}
\ethm

\bthm[\cite{panda}]
Given an $\faq$ $\varphi$ over the Boolean semiring with hypergraph $\calH=(\calV, \calE)$ and free variables $F \subseteq \calV$  over a database of size $N$, the $\panda$ algorithm can answer $\varphi$ in time
$O(|\calV|\cdot 2^{2^{|\calV|}}\cdot(\poly(\log N)\cdot N^{\subw_F(\calH)} + \log N\cdot \outputsize))$.
\label{thm:panda-runtime}
\ethm

%% file: example.tex
\section{Example}
\label{sec:example}

We illustrate our approach using the following Boolean query\footnote{If $R$, $S$, and $T$ would record direct train connections between cities, then this query would ask whether there exists a pair of cities with no direct train connection but with connections via another city.}:
\begin{equation}
C \leftarrow R(X,Y), S(Y,Z), \neg T(X,Z)
\label{eqn:Q:example}
\end{equation}
where all input relations have sizes
upper bounded by $N$ and thus the {\em active} domain of any variable $X$ has size at most $N$.
The query $C$ can be answered trivially in time $O(N^2)$ by joining
$R$ and $S$ first, and then, for each triple $(x,y,z)$ in the join, by verifying
whether $(x,z) \notin T$ with a (hash) lookup. Suppose we know that the degree of relation $T$
is less than two. Can we do better than $O(N^2)$ in that case? The answer is YES. 

\paragraph*{Rewriting to not-all-equal predicates}
By viewing $T$ as a bipartite graph of maximum degree two, 
it is easy to see that
$T$ can be written as a disjoint union of two relations $M_1(X,Z)$ and $M_2(X,Z)$
that represent {\em matchings} in the following sense:
for any $i \in [2]$, if $(x,z) \in M_i$ and $(x',z') \in M_i$, then either $(x,z)=(x',z')$
or $x \neq x'$ and $z \neq z'$. Let $\dom(Z)$ denote the active domain
 of the variable $Z$.
Define, for each $i \in [2]$, a singleton relation $W_i(Z) \leftarrow \dom(Z)\wedge \neg(\pi_ZM_i)(Z)$.
Clearly, $|W_i| \leq N$ and given $M_i$, $W_i$ can be computed in $O(N)$ preprocessing time.
For each $i \in [2]$, create a new variable $X_i$ with domain $\dom(X_i) = \dom(X)$. 
Then, 
\begin{align}
\neg M_i(X,Z) \equiv 
W_i(Z) \vee \exists X_i \bigl[ M_i(X_i,Z)  \wedge \nae(X, X_i) \bigr].\label{eqn:Ti}
\end{align}

The predicate $\nae$ stands for not-all-equal: It is the negation of the conjunction of pairwise equality on its variables. For arity two as in the rewriting of $\neg M_i(X,Z)$, $\nae(X, X_i)$ stands for the disequality $X\neq X_i$.

From $T = M_1 \vee M_2$ and~\eqref{eqn:Ti}, we can rewrite the
original query $C$ from~\eqref{eqn:Q:example} into a disjunction of Boolean conjunctive queries without negated relations but with one or two extra existential variables that are involved in {\em disequalities} ($\neq$): $C \equiv \bigvee_{i\in[4]} C_i$, where
\begin{align*}
C_1 \leftarrow & R(X,Y)\wedge S(Y,Z)\wedge W_1(Z)\wedge W_2(Z). \\
C_2 \leftarrow & R(X,Y)\wedge S(Y,Z)\wedge  W_1(Z)\wedge  M_2(X_2,Z)
                 \wedge X\neq X_2.  \\
C_3 \leftarrow & R(X,Y)\wedge S(Y,Z)\wedge  W_2(Z)\wedge  M_1(X_1,Z)
                 \wedge  X\neq X_1.  \\
C_4 \leftarrow & R(X,Y)\wedge S(Y,Z)\wedge  M_1(X_1,Z)\wedge  M_2(X_2,Z) 
                 \wedge X\neq X_1\wedge  X\neq X_2.
\end{align*}

It takes linear time to compute the matching
decomposition of $T$ into $M_1$ and $M_2$ since: (1) 
the relation $T$ is a bipartite graph with degree at most two, and
it is thus a union of even cycles and paths; and (2) we can traverse the cycles and paths
and add alternative edges to $M_1$ and $M_2$.
In general, when the maximum degree is higher and when $T$ is not a binary
predicate, Proposition~\ref{prop:matching:decomp} shows how to decompose a
relation into high-dimensional matchings efficiently. The number of queries $C_i$ depends exponentially on the arities and degrees of the negated relations.

\paragraph*{Boolean tensor decomposition} The acyclic query $C_1$ can be answered in $O(N\log N)$ time using for instance $\InsideOut$~\cite{faq}; this algorithm first sorts the input relations in time $O(N\log N)$. 
The query $C_2$ can be answered as follows.
Let $\forall i \in [\log N], f_i : \dom(X) \to \{0,1\}$ 
denote the function such that $f_i(X)$ is the $i$th bit of $X$ in its binary
representation.  Then, by noticing that
\begin{equation}
X \neq X_2 \equiv \bigvee_{b \in \{0,1\}}\bigvee_{i\in [\log N]}
f_i(X) = b \wedge f_i(X_2) \neq b
\label{eqn:boolean:tensor:decomp}
\end{equation}
we can break up the query $C_2$ into the disjunction of $2\log N$ acyclic queries 
of the form
\begin{multline}
C_2^{b,i} \leftarrow 
R(X,Y)\wedge S(Y,Z)\wedge  W_1(Z)\wedge  M_2(X_2,Z) 
\wedge f_i(X) = b\wedge  f_i(X_2) \neq b.
\end{multline}
For a fixed $b$, both $f_i(X)=b$ and $f_i(X_2)\neq b$ are singleton
relations on $X$ and $X_2$, respectively. Then, $C_2$ can be answered in time $O(N\log^2 N)$. The same applies to $C_3$. We can use the same trick to answer $C_4$ in time $O(N\log^3 N)$. However, we can do better than that by observing that 
when viewed as a Boolean tensor in~\eqref{eqn:boolean:tensor:decomp}, the disequality tensor has 
the Boolean rank bounded by $O(\log N)$.
In order to answer $C_4$ in time $O(N\log^2 N)$, we will show that the three-dimensio\-nal 
tensor $(X \neq X_1) \wedge (X \neq X_2)$ has the Boolean rank bounded
by $O(\log N)$ as well. To this end, we extend the color-coding 
technique. We can further shave off a $\log N$ factor in the complexities of $C_2$, $C_3$, and $C_4$, as explained in Section~\ref{sec:use-tensor-decomposition}.

\paragraph*{Construction of the Boolean tensor decomposition} 
We next explain how to compute a tensor decomposition for the conjunction of disequalities in $C_4$. 
We show that there exists a family $\calF$ of functions $f : \dom(X) \to \{0,1\}$
satisfying the following conditions:
\bi
 \item[(i)] $|{\cal F}| = O(\log |\dom(X)|) = O(\log N)$,
 \item[(ii)] For every triple $(x,x_1,x_2) \in \dom(X)^3$
 for which $x\neq x_1 \wedge x\neq x_2$, there is a function $f \in \calF$
 such that $f(x)\neq f(x_1) \wedge f(x) \neq f(x_2)$, and
 \item[(iii)] $\calF$ can be constructed in time $O(N\log N)$.
\ei
We think of each function $f$ as a ``coloring'' that assigns a
``color'' in $\{0,1\}$ to each element of $\dom(X)$.
Assuming $(i)$ to $(iii)$ hold, it follows that
\begin{multline}
X\neq X_1 \wedge X\neq X_2 \equiv 
\bigvee_{(c,c_1,c_2)}\bigvee_{f \in \calF}
f(X)=c \wedge f(X_1)=c_1 \wedge f(X_2)=c_2,
\end{multline}
where $(c,c_1,c_2)$ ranges over all triples in $\{0,1\}^3$ such that $c\neq c_1$
and $c \neq c_2$. Given this Boolean tensor decomposition, we can solve $C_4$ in time $O(N\log^2 N)$.

We prove $(i)$ to $(iii)$ using a combinatorial object
called the {\em disjunct matrices}.
These matrices are the central subject of combinatorial group
testing~\cite{NgoSurvey1999,MR1742957}.

\bdefn[$k$-disjunct matrix]
A $t \times N$ binary matrix $\bm A=(a_{ij})$ is called a {\em $k$-disjunct matrix}
if for every column $j \in [N]$
and every set $S \subseteq [N]$ such that $|S|\leq k$ and $j \notin S$, there exists
a row $i \in [t]$ for which $a_{ij}=1$ and $a_{ij'}=0$ for all $j' \in S$.
\edefn

It is known that for every integer $k< \sqrt N$, there exists a $k$-disjunct
matrix (or equivalently a combinatorial group testing~\cite{NgoSurvey1999}) with $t = O\left( k^2\log N\right)$ rows that  
can be constructed in time $O(k^2N\log N)$~\cite{DBLP:journals/tit/PoratR11}.
(If $k \geq \sqrt N$, we can just use the identity matrix.)
In particular, for $N = |\dom(X)|$ and $k=2$, 
a $2$-disjunct matrix $\bm A=(a_{ij})$ of size $O(\log N) \times N$ can be
constructed in time $O(N\log N)$. From the matrix we define the function family
$\calF$ by associating a function $f_i$ to each row $i$ of the matrix, and every
member $x \in \dom(X)$ to a distinct column $j_x$ of the matrix.
Define $f_i(x) = a_{i,j_x}$ and $(i)$--$(iii)$ straightforwardly follow.

%% file: results.tex
\section{Untangling bounded-degree relations}
\label{sec:results}

In this section we introduce a rewriting of queries of the form~\eqref{eqn:Q:R:F}
into queries with so-called {\em not-all-equal} predicates, under the assumption that the relation $R_S(\bm X_S)$ for every hyperedge $S \in \overline\calE$  has bounded degree $\deg(R_S)$.

\bdefn[Not-all-equal]
Let $k\geq 2$ be an integer, and $S$ be a set of $k$ integers. 
The relation $\nae_k(\bm X_S)$, or $\nae(\bm X_S)$ for simplicity, 
holds true iff not all variables in $\bm X_S$ are equal:
$\nae(\bm X_S) = \neg \bigwedge_{\{i,j\} \in \binom S 2} X_i=X_j.$
\edefn

The disequality ($\neq$) relation is exactly $\nae_2$. 
The negation of a matching is connected to $\nae$ predicates as follows.

\bprop
Let $M(\bm X_S)$ be a $k$-ary matching, where $k=|S|\geq 2$. For any $i,j \in S$,
define the unary relation $W_i(X_i) \leftarrow \dom(X_i)\wedge\neg (\pi_i M)(X_i)$ and the binary relation $M_{ij} = \pi_{i,j} M$. For any $\ell \in S$, it holds that
{\small
\begin{multline}
   \neg M(\bm X_S) \equiv \left(\bigvee_{i\in S\setminus \{\ell\}}
   W_i(X_i)\right) \vee \exists \bm Y_{S\setminus \{\ell\}} 
\left[
   \nae(X_{\ell}, \bm Y_{S\setminus \{\ell\}})
   \wedge 
   \bigwedge_{j \in S\setminus \{\ell\}} M_{\ell j}(Y_j,X_j) 
\right].
\label{eqn:matching:nae}
\end{multline}
}
\label{prop:matching:nae}
\eprop
\bp
The intuition for this rewriting is as follows. A value $x_i\in \dom(X_i)$ occurs in at most one tuple in the matching $M$. Therefore, any value in a tuple determines the rest of the tuple. The rewriting in~\eqref{eqn:matching:nae} first turns every tuple in $M$ into a tuple whose values are all the same, i.e., all-equal values. The negation of $M$ consists of tuples  with at least two different values, i.e., not-all-equal values. 

We next prove that the rewriting is correct.

In one direction, consider a tuple $\bm x_S \notin M$, 
i.e., $\neg M(\bm x_S)$ holds, and suppose $x_i \notin W_i$ for all $i \in S
\setminus \{\ell\}$.
This means, for every $i \in S\setminus\{\ell\}$, there is a unique tuple $\bm t^{(i)} =
(t^{(i)}_j)_{j \in S} \in M$ such that $x_i = t^{(i)}_i$. 
Define $y_j = t^{(j)}_{\ell}$ for all $j \in
S\setminus\{\ell\}$. The tuple $\bm y_{S\setminus \{\ell\}}$ satisfies 
$(y_j,x_j) = (t^{(j)}_{\ell}, t^{(j)}_j) \in M_{\ell j}$, for all $j \in
S\setminus \{\ell\}$. Moreover, one can verify that $\nae(x_{\ell},\bm y_{S\setminus \{\ell\}})$
holds.
In particular, if $y_j=x_\ell$ for all $j \in S\setminus\{\ell\}$, then all tuples $\bm t^{(j)}\in M$ are the same tuple (since $M$ is a matching) and that tuple is $\bm x_S$. Hence $\bm x_S\in M$ which is a contradiction.

Conversely, suppose there exists a tuple $(\bm x_S, \bm y_{S \setminus
\{\ell\}})$ satisfying the right hand side of~\eqref{eqn:matching:nae}. 
If $x_i \in W_i$ for any $i \in S\setminus\{\ell\}$, then $\bm
x_S \notin M$, i.e., $\bm x_S$ satisfies the left hand side
of~\eqref{eqn:matching:nae}. Now, suppose $x_i \notin W_i$ for all $i \in
S\setminus\{\ell\}$. Suppose to the contrary that $\bm x_S \in M$. Then, for all $j\in S\setminus \{\ell\}$ we have $y_j=x_{\ell}$ since $M_{\ell j}(y_j,x_j)$ must hold. This means that 
$\nae(x_{\ell},\bm y_{S\setminus\{\ell\}}) = \neg\wedge_{j\in S\setminus \{\ell\}} x_{\ell} = y_j$ does not hold. This contradicts our hypothesis.
\ep

We use the connection to $\nae$ predicates to decompose a query containing a
conjunction of negated boun\-ded-degree relations into a disjunction of positive terms, as given next by Proposition~\ref{prop:reduced:form}. We call this rewriting {\em untangling}.

Let $\fhtw_F$ and $\subw_F$ denote the fractional hypertree width and respectively the submodular width of the conjunctive query $Q(\bm X_F) \leftarrow 
   \skeleton$ (These notions are defined in Section~\ref{subsec:insideout-panda}). 

\bprop
Let $Q$ be the query defined in Eq.~\eqref{eqn:Q:R:F}:
$Q(\bm X_F) \leftarrow 
   \skeleton \wedge \bigwedge_{S\in\overline\calE} \neg R_S(\bm X_S)$.
We can compute in linear time a collection of $B$ hypergraphs $\calH_i = (\calV_i,\calE_i)$ such that
\begin{align}
   Q(\bm X_F) \equiv \bigvee_{i \in [B]} Q_i(\bm X_F), && \text{ where } \forall i
   \in [B]: Q_i(\bm X_F) \leftarrow \skeleton_i \wedge \bigwedge_{S \in \calE_i} \nae(\bm Z_S),
\label{eqn:reduced:form}
\end{align}
and $\skeleton_i$ is the body of a conjunctive query satisfying
\begin{align*}
   \fhtw_F(\skeleton_i) &\leq \fhtw_F(\skeleton), &\text{ and } &&
   \subw_F(\skeleton_i) \leq \subw_F(\skeleton).
\end{align*}
Furthermore, the number $B$ of queries is bounded by
$B \leq \prod_{S \in \overline\calE} (|S|)^{|S|(\deg(R_S)-1)+1}$.
\label{prop:reduced:form}
\eprop
\bp
From Proposition~\ref{prop:matching:decomp}, each relation $R_S(\bm X_S)$ can be
written as a disjoint union of $D_S \leq |S|(\deg(R_S)-1)+1$ matchings $M_S^\ell$, $\ell \in
[D_S]$. These matchings can be computed in linear time. 
Hence, the
second half of the body of query $Q$ can be rewritten equivalently as
\[
   \bigwedge_{S\in\overline\calE} \neg R_S(\bm X_S)
   \equiv 
   \bigwedge_{S\in\overline\calE} \neg \bigvee_{\ell \in [D_S]} M_S^\ell (\bm X_S)
   \equiv \bigwedge_{\substack{S\in\overline\calE\\ \ell  \in [D_S]}} \neg
   M_S^\ell (\bm X_S).
\]
To simplify notation, let $\overline\calE_1$ denote the multiset of edges obtained
from $\overline\calE$ by duplicating the edge $S \in \overline\calE$ exactly $D_S$ times.
Furthermore, for the $\ell$-th copy of $S$, associate the matching $M^\ell_S$
with the copy of $S$ in $\overline\calE_1$; use $M_S$ to denote the matching
corresponding to that copy. Then, we can write $Q$ equivalently
$ Q(\bm X_F) \leftarrow \skeleton \wedge \bigwedge_{S\in \overline\calE_1} \neg M_S(\bm X_S). $

For each $S \in \overline\calE_1$, fix an arbitrary integer $\ell_S \in S$.
From Proposition~\ref{prop:matching:nae}, the negation of $M_S$ can be written
as
{\small
\begin{multline*}
   \neg M_S(\bm X_S) \equiv
   \biggl(\bigvee_{i \in S \setminus \{\ell_S\}} W_{i}^{S}(X_i)\biggr) \vee \exists \bm Y^{S}_{S\setminus \{\ell_S\}} 
\biggl[
   \bigwedge_{j \in S\setminus \{\ell_S\}} (\pi_{\ell_S,j} M_S)(Y^{S}_j,X_j) 
   \wedge \nae(X_{\ell_S}, \bm Y^{S}_{S\setminus \{\ell_S\}})
\biggr],
\end{multline*}
}
where $W^{S}_i$ is a unary relation on variable $X_i$,
and $\bm Y^{S}_{S\setminus \{\ell_S\}}=(Y^{S}_i)_{i \in S\setminus
\{\ell_S\}}$ is a tuple of fresh variables, only associated with (the copy of) $S$. 
In particular, if $S$ and $S'$ are two distinct items in the multiset $\overline\calE_1$, then $Y^{S}_i$ and $Y^{S'}_i$ are two distinct variables.

Each negated term $\neg M_S(\bm X_S)$ is thus expressed as a disjunction of $|S|$ positive terms. We can then express the conjunction of $|\overline\calE_1|$ negated terms as the disjunction of $\prod_{S\in \overline\calE_1} |S|$ conjunctions. For this, define a collection of tuples $\calT = \prod_{S \in \overline\calE_1} S$. In particular, every member $\bm T \in \calT$ is a tuple $\bm T = (t_S)_{S\in\overline\calE_1}$ where $t_S \in S$.
The second half of the body of query $Q$ can be rewritten 
equivalently as
\begin{align*}
   &\bigwedge_{S\in\overline\calE} \neg R_S(\bm X_S)
   \equiv 
   \bigwedge_{S\in\overline\calE_1} \neg M_S (\bm X_S)\\
   &\hspace*{-1.25em}\equiv 
   \bigwedge_{S \in \overline\calE_1} \left( \bigvee_{i \in S\setminus\{\ell_S\}}W_i^{S}(X_i) \vee 
   \exists \bm Y^{S}_{S\setminus \{\ell_S\}}\right.
   \left. \bigl[ \bigwedge_{j \in S\setminus\{\ell_S\}}
   (\pi_{\ell_S,j}M_S)(Y^{S}_j,X_j) \wedge \nae\bigl(X_{\ell_S},  \bm Y^{S}_{S\setminus \{\ell_S\}} 
   \bigr) \bigr]
   \right)\\
   &\hspace*{-1.25em}\equiv
   \bigvee_{\bm T \in \calT}
   \bigwedge_{\substack{S \in \overline\calE_1\\ t_S \neq \ell_S}}
   W_{t_S}^S(X_{t_S}) \wedge 
   \bigwedge_{\substack{S \in \overline\calE_1\\ t_S = \ell_S}}
   \exists \bm Y^{S}_{S\setminus \{\ell_S\}} 
     \bigl[ \bigwedge_{j \in S\setminus\{\ell_S\}}
   (\pi_{\ell_S,j} M_S)(Y^{S}_j,X_j) \wedge \nae\bigl(X_{\ell_S},  \bm Y^{S}_{S\setminus \{\ell_S\}} 
   \bigr) \bigr]
\end{align*}

The original query $Q$ is equivalent to the
disjunction
\[ Q(\bm X_F) \equiv \bigvee_{\bm T \in \calT} Q_{\bm T}(\bm X_F) \]
of up to $\prod_{S\in \overline\calE_1} |S|$ queries $Q_{\bm T}$ defined by
\begin{multline}
    \underbrace{\skeleton \wedge
   \bigwedge_{\substack{S \in \overline\calE_1\\ t_S \neq \ell_S}}
   W_{t_S}^S(X_{t_S}) \wedge 
   \bigwedge_{\substack{S \in \overline\calE_1\\ t_S = \ell_S\\ j \in S\setminus \{\ell_S\}}}
   (\pi_{\ell_S,j}M_S)(Y^{S}_j,X_j)}_{\skeleton_i}  
   \wedge \bigwedge_{\substack{S \in \overline\calE_1\\ t_S = \ell_S}}
   \nae\bigl(X_{\ell_S},  \bm Y^{S}_{S\setminus \{\ell_S\}}
   \bigr)
   \label{eqn:reduced:form:detailed}
\end{multline}

In the above definition of $Q_{\bm T}$, let us denote all but the last conjunction of $\nae$ predicates by $\skeleton_{i}$.
It holds that
$\fhtw_F(\skeleton_i) \leq \fhtw_F(\skeleton)$, and
$\subw_F(\skeleton_i) \leq \subw_F(\skeleton)$ by Lemma~\ref{lmm:subw_i<=subw}.
We now turn to the conjunction of $\nae$ predicates in \eqref{eqn:reduced:form:detailed}.
Since each $S \in \overline\calE$ is repeated at most $|S|(\deg(R_S)-1)+1$ times in $\overline\calE_1$, it follows that the number $\prod_{S\in \overline\calE_1} |S|$ of conjunctive queries $Q_{\bm T}$ is at most $\prod_{S\in \overline\calE} |S|^{|S|(\deg(R_S)-1)+1}$.
\ep

\section{Boolean tensor decomposition}
\label{subsec:boolean:tensor:decomp}

Thanks to the untangling result in Proposition~\ref{prop:reduced:form}, we only need to concentrate on answering queries of the
form~\eqref{eqn:reduced:form}.
To deal with the conjunction of $\nae$ predicates,
this section describes the construction of a Boolean tensor decomposition of a conjunction $\bigwedge_{S\in \calA} \nae(\bm X_S)$ of $\nae$ predicates. 
The multi-hypergraph of this conjunction has the query variables as vertices and the $\nae$ predicates as hyperedges.

\blmm\label{lemma:boolean-tensor-decomposition}
Let $\calG = (U, \calA)$ be the multi-hypergraph of a conjunction $\bigwedge_{S\in \calA} \nae(\bm X_S)$, $N$ an upper bound on the domain sizes for variables
$(X_i)_{i \in U}$, and $c$ a positive integer.
Suppose there exists a family $\calF$ of functions $f : [N] \to [c]$ satisfying
the following property
\begin{align}
& \text{for any proper $N$-coloring $h : U \to [N]$ of $\calG$
there exists a function $f \in \calF$} \label{eqn:good:condition}\\
& \text{such that $f \circ h$ is a proper $c$-coloring of $\calG$.}\nonumber
\end{align}
Then, the following holds:
\begin{equation}
\bigwedge_{S\in \calA} \nae(\bm X_S) \equiv
\bigvee_{g} \bigvee_{f\in \calF} \bigwedge_{i \in U} f(X_i) = g(i),
\label{eqn:decomp:id}
\end{equation}
where $g$ ranges over all proper $c$-colorings of $\calG$. In particular, the
Boolean tensor rank of the left-hand side of~\eqref{eqn:decomp:id} is bounded by 
$r = P(\calG, c) \cdot |\calF|$.
\label{lmm:decomp:id}
\elmm
\bp
Let $\bm x_U$ denote any tuple satisfying the LHS of~\eqref{eqn:decomp:id}.
Define $h : U \to [N]$ by setting $h(i) = x_i$. Then $h$ is a proper $N$-coloring
of $\calG$, which means there exists $f \in \calF$ such that $g = f\circ h$ is
a proper $c$-coloring of $\calG$. Then the conjunct on the RHS corresponding
to this particular pair $(g,f)$ is satisfied.

Conversely, let $\bm x_U$ denote any tuple satisfying the RHS of~\eqref{eqn:decomp:id}.
Then, there is a pair $(g,f)$ whose corresponding conjunct on the RHS
of~\eqref{eqn:decomp:id} is satisfied, i.e., $f(x_i) = g(i)$ for all $i \in U$. 
Recall that $g$ is a proper $c$-coloring of $\calG$. 
If there exists $S \in \calA$ such that $\nae(\bm x_S)$ does {\em not} hold, then $x_i=x_j$
for all $i,j\in S$, implying $g(i) = f(x_i) = f(x_j) = g(j)$ for all $i,j\in S$,
contradicting the fact that $g$ is a proper coloring.

For the Boolean tensor rank statement,
note that~\eqref{eqn:decomp:id} is a Boolean tensor decomposition of the formula
$\bigwedge_{S\in\cal A} \nae(\bm X_S)$, because $f(X_i) = g(i)$ is a unary
predicate on variable $X_i$. This predicate is of size bounded
by $N$.
\ep
To explain how Lemma~\ref{lmm:decomp:id} can be applied, we exemplify two
techniques, showing the intimate connections of our Boolean tensor decomposition problem to combinatorial group testing and perfect hashing.

\begin{ex}[Connection to group testing]
   \label{ex:group:testing}
   Consider the case when the graph $\calG$ is a {\em $k$-star}, i.e., a tree with a center
   vertex and $k$ leaf vertices.
   Let $\bm A$ be a $O(k^2\log N) \times N$ binary $k$-disjunct matrix, which
   can be constructed in time $O(kN\log N)$ (This is due to known results on $k$-restriction and error codes, recalled in Appendix~\ref{subsec:prelim:k-restrict}). We can
   assume $k< \sqrt N$ to avoid triviality. 
   Consider a family $\calF$ of functions $f : [N] \to \{0,1\}$ constructed as
   follows: there is a function $f$ for every row $i$ of $\bm A$, where $f(j) =
   a_{ij}$, for all $j \in [N]$. 
   The family $\calF$ has size $O(k^2\log N)$. We show that $\calF$ satisfies
   condition~\eqref{eqn:good:condition}.
   Let $h : U \to [N]$ denote any coloring of the
   star. Let $j \in [N]$ be the color $h$ assigns to the center, and $S$ be the
   set of colors assigned to the leaf nodes. Clearly $j \notin S$. Hence, there
   is a function $f \in \calF$ for which $f(j) =1$ and $f(j')=0$ for all $j'\in
   S$, implying $f \circ h$ is a proper $2$-coloring of $\calG$. 

   A consequence of our observation is that for a $k$-star $\calG$ the conjunction $\bigwedge_{S\in\cal A} \nae(\bm X_S)$ has Boolean
   rank bounded by $O(k^2\log N)$.
   \nop{This is a generalization of~\eqref{eqn:boolean:tensor:decomp}.}
   \qed
\end{ex}

\begin{ex}[Connection to perfect hashing]
   Consider now the case when the graph $\calG$ is a $k$-clique.
   Let $\calF$ denote any $(N,c,k)$-perfect hash family, i.e., a family of hash
   functions from $[N] \to [c]$ such that for every subset $S\subseteq [N]$ of
   size $k$, there is a function $f$ in the family for which its image is also
   of size $k$.
   It is easy to see that this hash family satisfies~\eqref{eqn:good:condition}.
   From~\cite{DBLP:journals/talg/AlonMS06}, it is known that we can construct in 
   polytime an $(N,k^2,k)$-perfect hash family of size $O(k^4\log N)$.
   However, it is not clear what the runtime exponent of their construction is.
   What we need for our application is that the construction should run in
   linear data complexity and in polynomial query complexity. We use a result 
   from~\cite{DBLP:journals/tit/PoratR11} to exhibit such a construction in
   Theorem~\ref{thm:prob:const};
   furthermore, our hash family has size only $O(k^2\log N)$.
   \qed
\end{ex}

We next construct the smallest family $\calF$ satisfying Lemma~\ref{lemma:boolean-tensor-decomposition}. We first
bound the size of $\calF$ using the probabilistic method~\cite{AlonSpencer:probabilistic} and then specify how to
derandomize the probabilistic construction of $\calF$ to obtain a deterministic algorithm. For this, we need some terminology.

Every coloring $h : U \to [N]$ of $\calG=(U,\calA)$ induces a homomorphic image
$h(\calG) = (h(U), h(\calA))$, which is the graph on vertex set 
$h(U)$ and edge set $h(\calA)$ defined by
\begin{align*}
   h(U) = \{ h(v) \suchthat v \in U \} \subseteq [N],\quad\quad\quad
   h(\calA) = \bigl\{ h(S)=\{h(v) \suchthat v \in S\} \suchthat S \in \calA \bigr\} \subseteq 2^{[N]}.
\end{align*}
Here, we overload notation to allow $h$ range over sets and graphs.
Let $\col(\calG,N)$ denote the set of proper $N$-colorings $h$ of $\calG$. Each such proper $N$-coloring is a homomorphic image of $\calG$. Define $c$ as the maximum chromatic number over all homomorphic images of $\calG$: $c =\max_{h \in \col(\calG,N)}\chi(h(\calG))$.
For a given $h \in \col(\calG,N)$,
let $g : h(U) \to [c]$ be a proper $c$-coloring of $h(\calG)$. The {\em
multiplicity} of a color $i \in [c]$ is the number of vertices colored $i$ by
$g$. The {\em signature} of $g$ is the vector $\bm \mu(g) = (\mu_i)_{i\in
[c]}$, where $\mu_i$ is the multiplicity of color $i$.
Let $T_c(h)$ denote the collection of all signatures of proper $c$-colorings of 
$h(\calG)$. For a given signature $\bm \mu =(\mu_1,\dots,\mu_c) \in T_c(h)$, 
let $n(\bm \mu, h)$ denote the number of
proper $c$-colorings of $h(\calG)$ whose signature is $\bm \mu$.

\begin{ex}
Suppose $\calG$ is the $k$-clique and $c=k$. Then, every
proper $k$-coloring of $h(\calG)$ has signature $\bm\mu = \bm 1_k = (1,1,\ldots,1)$: 
$T_c(h)$ has only one member, but $n(\bm 1, h) = k!$.
If $\calG$ is the $k$-star then $c=2$ and for any $h \in \col(\calG, N)$,
$h(\calG)$ is an $\ell$-star for some $\ell \in [k]$. Then, 
$T_2(h)$ has two signatures: $\bm \mu=(\ell, 1)$ and $\bm \mu'=(1,\ell)$; 
furthermore, $n(\bm \mu,h)=n(\bm \mu',h)=1$.\qed
\end{ex}

\bdefn[Strongly Explicit Construction]
A family $\calF$ of functions $f : [N] \to [c]$ is 
said to be {\em strongly explicit} if there is an
algorithm that, given an index to a function $f$ in $\calF$ and a number $j \in [N]$, 
returns $f(j)$ in $\poly(\log |\calF|,\log N)$-time. 
\edefn

The next theorem gives two upper bounds on the size of a family of hash functions satisfying~\eqref{eqn:good:condition} that we use to define the rank of our Boolean tensor decomposition: The first bound is for such families in general, whereas the second is for strongly explicit families that we can use effectively.

\begin{thm}
Let $\calG = (U, \calA)$ be a multi-hypergraph,
 $c = \max_{h \in \col(\calG,N)}\chi(h(\calG))$, and
 $\bm p =(p_1,\dots,p_c) \in \R_+^c$ be a fixed
   non-negative real vector such that $\norm{\bm p}_1=1$.
   Define
   \begin{equation}
      \theta(\bm p) = \min_{h \in \col(\calG,N)} 
      \sum_{\bm \mu\in T_c(h)} n(\bm \mu,h) \prod_{i=1}^c p_i^{\mu_i}
      \label{eqn:theta:p}
   \end{equation}
   Then, the following hold:
   
   \bi
\item[(a)] There exists a family $\calF$ of functions $f : [N] \to [c]$ satisfying
   \eqref{eqn:good:condition} such that
\begin{equation}
   |\calF| \leq \left\lceil \frac{\ln P(\calG,N)}{\theta(\bm p)} \right\rceil
   \leq
   \frac{|U| \log N}{\theta(\bm p)}.
   \label{eqn:F:upper:bound}
\end{equation}
   \label{thm:chromatic}
\item[(b)] There is a strongly explicit family $\calF'$ of functions $f : [N] \to [c]$ 
   satisfying~\eqref{eqn:good:condition} such that
\begin{equation}
   |\calF'| = O\left( \frac{|U|^3 \cdot \log |U| \cdot \log N}
                           {\theta(\bm p)}
               \right).
\end{equation}
   \ei
   \label{thm:prob:const}
\end{thm}

The next corollary follows immediately from Lemma~\ref{lmm:decomp:id} and Theorem~\ref{thm:prob:const}.
\bcor
Let $\calG = (U,\calA)$ be a multi-hypergraph, $c = \max_{h \in
\col(\calH,N)} \chi(h(\calG))$, and
\begin{equation}
   \theta^* = \max_{\bm p : \norm{\bm p}_1=1, \bm p \geq \bm 0} \theta(\bm p),
\end{equation}
where $\theta(\bm p)$ is defined in~\eqref{eqn:theta:p}. 
The following hold:
\bi
\item[(a)]
The Boolean rank of the function $\bigwedge_{F\in\calA}\nae(\bm X_F)$ 
is upper bounded by $\frac{P(\calG, c) \cdot \ln P(\calG, N)}{\theta^*}$. 
\item[(b)]
Given $\bm p$, there is a strongly explicit Boolean tensor 
decomposition of $\bigwedge_{F\in\calA}\nae(\bm X_F)$ whose rank is upper bounded by 
$P(\calG,c) \cdot \frac{|U|^3 \cdot \log |U| \cdot \log N}{\theta(\bm
p)}.$
\ei
\label{cor:bounding-boolean-rank}
\ecor

To apply the above result, we need to specify $\bm p$ to maximize $\theta(\bm p)$.
We do not know how to compute the optimizer $\bm p^*$ in closed form. We next
discuss several observations that allow us to bound
$\theta^*$ from below or compute it exactly.
In the following, for any tuple $\bm \mu = (\mu_1,\dots,\mu_\ell)$ of positive
integers, let $K_{\bm \mu}$ denote the complete $\ell$-partite
graph defined as follows. For every $i \in [\ell]$ there is an independent set $I_i$ of size $\mu_i$.
All independent sets are disjoint. The vertex set is $\bigcup_{i\in[\ell]} I_i$ and the
vertices not belonging to the same independent set are connected.
Without loss of generality, we assume $\mu_1\geq \cdots \geq \mu_\ell$ when specifying the graph
$K_{\bm \mu}$. For example, $K_{\bm 1_k}$ is the $k$-clique, and $K_{(k,1)}$ is
the $k$-star.

\begin{prop}
   \label{prop:theta:lower:bound}
The following hold:
\bi
\item[(a)] Given a multi-hypergraph $\calG=(U,\calA)$ with $|U|=k$ and $c = \max_{h \in
   \col(\calG,N)}\chi(h(\calG))$, it holds that
   $\theta^* \geq \frac{1}{c^c} \geq \frac{1}{k^k}$.
\item[(b)] Suppose $\calG = K_{\bm \mu}$ for some positive integer tuple $\bm
   \mu = (\mu_1,\dots,\mu_\ell)$, where $\mu_1\geq \cdots \geq \mu_\ell \geq 1$. 
   Let $S_\ell$ denote the set of all permutations of $[\ell]$, and 
   $\fp(\bm \mu)$ denote the number of permutations $\pi \in S_\ell$ for which
   $\mu_i = \mu_{\pi(i)},\forall i \in [\ell]$. Then,
   \begin{align}
      \theta^* = \max_{\bm p} \sum_{\pi \in S_\ell} \prod_{i=1}^\ell
      p_i^{\mu_{\pi(i)}} 
      \geq \sum_{\pi\in S_\ell} \prod_{i=1}^\ell \left(\frac{\mu_i}{\norm{\bm
      \mu}_1}\right)^{\mu_{\pi(i)}} 
      \geq \fp(\bm \mu) \prod_{i=1}^\ell \left(\frac{\mu_i}{\norm{\bm
      \mu}_1}\right)^{\mu_{i}}.\label{eqn:theta*:combined}
   \end{align}
\ei
\end{prop}

\bcor
Let $\ell \in [k]$ be an integer. Let $\bm \mu = (k-\ell, \bm
1_\ell)$. Then, when $\calG = K_{\bm \mu}$ we have 
\[ \theta^* \geq
\frac{\ell!}{k^\ell} \left(\frac{k-\ell}{k}\right)^{k-\ell}
\geq \frac{\ell!}{e^\ell} \frac{1}{k^\ell},
\]
where $e=2.7..$ is the base of the natural log.
In particular, $\calG$ is a $(k-1)$-star when $\ell = 1$ and the bound is
$\theta^* \geq \frac{1}{ek}$. When $\ell=k$, then $\calG$ is a $k$-clique
and the bound is $\theta^*=k!/k^k$.
\ecor
For any constant $\ell\in[k]$, the bound for $\theta^*$ is $\Omega(1/k^\ell)$; in
particular, the lower bound for $\theta^*$ ranges anywhere between
$\Omega(1/k)$, $\Omega(1/k^2)$, up to $\Omega(k!/k^k)$. There is a spectrum of
these bounds, leading to a spectrum of Boolean tensor ranks for our decomposition.

\begin{ex}
From~\eqref{eqn:F:upper:bound} and the above corollary, it follows
that when $\calG$ is a $k$-star, the corresponding Boolean rank is bounded by
$O(k^2\log N)$, matching the group testing connection from
Example~\ref{ex:group:testing}. The reason is twofold. We need two colors to color a $k$-star and the chromatic polynomial of a $k$-star using two colors is two. The size of the family $\calF$ of hash functions is upper bounded by $\frac{|U| \log N}{\theta^*}$ where $\theta^*$ is at least $\frac{1}{ek}$ and $|U|=k+1$. Then, $|\calF|\leq e\cdot k\cdot (k+1)\log N  = O(k^2\log N)$. This matches the
tailor-made construction from Example~\ref{ex:group:testing}.
However, our strongly explicit construction in Theorem~\ref{thm:prob:const}(b) 
yields a slightly larger Boolean tensor decomposition of rank $O(k^4\log k\log N)$.

When applying part (b) of Proposition~\ref{prop:theta:lower:bound} to the problem of detecting 
$k$-paths in a graph, i.e., the query $P$ in the introduction, we obtain the Boolean rank $O(\frac{k^{k+3}}{k!}\cdot\log k\cdot\log N)$. This is because (1) we would need two colors and the chromatic polynomial for the $k$-path hypergraph using two colors is two,  and (2) the size of the family of strongly explicit functions is $O(\frac{(k+1)^3\log (k+1)\log N}{\theta^*})$ with $\theta^*=k!/k^k$.
\nop{our bound yields the result of Plehn and Voigt~\cite{PlehnV90}.}\qed
\end{ex}

\section{How to use the tensor decomposition}
\label{sec:use-tensor-decomposition}

Sections~\ref{sec:results} and ~\ref{subsec:boolean:tensor:decomp} introduced two rewriting steps. The first step transforms a conjunctive query with negation of the form~\eqref{eqn:Q:R:F} into a disjunction of conjunctive queries with $\nae$ predicates of the form~\eqref{eqn:reduced:form}. The second step transforms a conjunction of $\nae$ predicates into a disjunction of conjunctions of one-variable-conditions of the form~\eqref{eqn:decomp:id}. The first step exploited the bounded degrees of the negated relations to bound from above the number of disjuncts and independently of the database size. The second step uses a generalization of the color-coding technique to further rewrite a conjunction of $\nae$ predicates into a Boolean tensor decomposition whose rank depends on the structure of the multi-hypergraph of the conjunction. 
Both rewriting steps preserve the equivalence of the queries. 

In this section, we show that the query obtained after the two rewriting steps can be evaluated efficiently. This query has the form $Q(\bm X_F) \leftarrow \bigvee_{j \in [B]} Q_j(\bm X_F)$ where $\forall j \in [B]$:
\begin{equation}
   Q_j(\bm X_F) \leftarrow \bigvee_{g\in\col(\calG_j,c_j)} \bigvee_{f\in\calF_j} \underbrace{\left[\bigwedge_{S\in\calE_j} R_S(\bm
   X_S) \wedge \bigwedge_{i\in U_j} f(X_i)=g(i)\right]}_{Q_j^{(g, f)}(\bm X_F)}
   \label{eqn:prototypical}
\end{equation}
In particular, we will show that the data complexity of any conjunctive query with negation of the form~\eqref{eqn:Q:R:F} is the same as for its positive subquery $Q(\bm X_F) \leftarrow \skeleton$.

The subsequent development in this section uses the $\InsideOut$ algorithm and the $\faq$ framework (see Section~\ref{subsec:insideout-panda} and~\cite{faq}). 
For each $j \in [B]$, we distinguish two multi-hypergraphs for the query $Q_j(\bm X_F)$: $\calH_j = (V_j,\calE_j)$ and associated relations $(R_S)_{S\in\calE_j}$ for $\bigwedge_{S\in\calE_j} R_S(\bm X_S)$; and $\calG_j = (U_j, \calA_j)$ for $\bigwedge_{i\in U_j} f(X_i)=g(i)$, where $U_j\subseteq V_j$.
For the rest of this section, we will fix some $j \in [B]$ and drop the subscript $j$ for brevity. In particular, we will use $\calH=(V, \calE), \calG=(U, \calA), \calF$ to denote
$\calH_j=(V_j, \calE_j), \calG_j=(U_j, \calA_j), \calF_j$ respectively.

\paragraph*{A better semiring for shaving off a $\log N$ factor}
Let $r = P(\calG,c) \cdot |\calF|$ denote the Boolean tensor rank in the
decomposition~\eqref{eqn:decomp:id}.
If we were only interested in bounding the rank, we can use the bound on $|\calF|$ from Part (a) of Theorem~\ref{thm:prob:const}.
However, for the purpose of using the Boolean tensor decomposition in an algorithm, we have to be able to explicitly and efficiently construct the family $\calF$ of functions. We thus need to use the bound on $|\calF|$ from Part (b) of Theorem~\ref{thm:prob:const}.
To facilitate the explanations below, define $w = |\calF|/\log N$ so that
the Boolean rank is decomposed into $r = P(\calG,c) \cdot w \cdot \log N$;
that is, $w=\frac{|U|^3 \cdot \log |U|}{\theta(\bm p)}$ from Part (b) of Theorem~\ref{thm:prob:const}.

By Theorem~\ref{thm:insideout-runtime}, we can answer query~\eqref{eqn:prototypical} by running $r$ 
instantiations of $\InsideOut$, each of which computes $Q_j^{(g, f)}$ for some fixed pair $(g, f)$, and then take the disjunction of $Q_j^{(g, f)}$ over $g$ and $f$. The runtime is
\begin{equation} O(P(\calG,c) \cdot w \cdot (|\calE|+|U|)\cdot|V|^2\cdot (\log
N)^2\cdot(N^{\fhtw_F(\calH)}+\outputsize) ).
\end{equation}
The atoms $f(X_i)=g(i)$ are singleton factors, i.e., factors on one variable, and thus
do not increase the fractional hypertree width or the submodular width of the query.

These $r$ instantiations of $\InsideOut$ are run on sum-product instances over 
the Boolean semiring.
We can however reformulate the problem as sum-product over a different 
semiring, which
helps reduce the runtime. The new semiring $(\bm D, \oplus, \otimes, \bm 0, \bm
1)$ is defined as follows. The domain $\bm D$ is set to $\bm D = \{0,1\}^r$, the
collection of all $r$-bit vectors. The ``addition'' and ``multiplication''
operators $\oplus$ and $\otimes$ are bit-wise $\max$ and $\min$ (essentially,
bit-wise $\vee$ and $\wedge$). The additive identity is $\bm 0 = \bm 0_r$, the
$r$-bit all-$0$ vector. The multiplicative identity is $\bm 1 = \bm 1_r$, the
$r$-bit all-$1$ vector. To each input relation $R_S$, we associate a function
$\psi_S(\bm X_S) : \prod_{i\in S}\dom(X_i) \to \bm D$, where
$\psi_S(\bm x_S) = \bm 1$ if $\bm x_S \in R_S$ and $\bm 0$ otherwise.
Also, define $|U|$ extra singleton factors 
$\overline \psi_i : \dom(X_i) \to \bm D\ (\forall i \in U)$, where 
\begin{align}
   \overline \psi_i(x_i) = (b_{g,f})_{g \in \col(\calG,c), f \in \calF},\quad\text{where}\quad
   b_{g,f} &= 
\begin{cases}
   1 & \text{ if }f(x_i) = g(i)\\
   0 & \text{ if }f(x_i) \neq g(i).
\end{cases}
\end{align}
\bprop
The query~\eqref{eqn:prototypical} is equivalent to the
following $\sumprod$ expression
\begin{equation}
   \varphi(\bm x_F) = \bigoplus_{x_{|F|+1}}\cdots\bigoplus_{x_{|V|}}
   \bigotimes_{S\in\calE}\psi_S(\bm x_S) \otimes 
   \bigotimes_{i\in U}\overline\psi_i(x_i).
\end{equation}
The runtime of $\InsideOut$ for the expression $\varphi(\bm x_F)$ is 
\begin{equation}
O(P(\calG,c) \cdot w \cdot (|\calE|+|U|)\cdot|V|^2\cdot\log N\cdot (N^{\fhtw_F(\calH)}+\outputsize)).
   \label{eqn:after:idea:1}
\end{equation}
\eprop
\bp
For any $\bm x_F$, we have $Q(\bm x_F) = \true$ iff $\varphi(\bm x_F)$ $\neq \bm 0$.
This is because for each $\bm x_F$, the value $\varphi(\bm x_F)\in\D$ is an $r$-bit vector where each bit represents the answer to $Q_j^{(g, f)}(\bm X_F)$ for some pair $(g, f)$ (There are exactly $r = P(\calG,c) \cdot |\calF|$ such pairs).

The runtime of $\InsideOut$ follows from the observation that each operation
$\oplus$ or $\otimes$ can be done in $O(r/\log N)$-time, because those are
bit-wise $\vee$ or $\wedge$ and the $r$-bit vector can be stored in $O(r/\log N)$
words in memory. Bit-wise $\vee$ and $\wedge$ of two words are done in
$O(1)$-time.
\ep

We can further lower the data complexity of our approach using $\panda$ (See Section~\ref{subsec:insideout-panda} and~\cite{panda}).
By Theorem~\ref{thm:panda-runtime}, the complexity from \eqref{eqn:after:idea:1} becomes: 
\begin{equation}
O(P(\calG,c) \cdot w \cdot |V|\cdot 2^{2^{|V|}}\cdot(\poly(\log N)\cdot N^{\subw_F(\calH)}+\log N\cdot\outputsize)).
   \label{eqn:panda:final}
\end{equation}

\paragraph*{The data complexity for conjunctive queries with negation}

We are now ready to prove Theorem~\ref{thm:complexity}.
From Proposition~\ref{prop:reduced:form}, we untangle $Q$ into a disjunction of $B$ different queries $Q_j$ for $j \in [B]$ of the form~\eqref{eqn:reduced:form} where
$B \leq \prod_{S \in \overline\calE} (|S|)^{|S|(d_S-1)+1}=O(1)$
in data complexity.
From~\eqref{eqn:decomp:id}, each of these queries is equivalent to query~\eqref{eqn:prototypical}.
For a fixed ${g\in\col(\calG,c)}$ and ${f\in\calF}$, the inner conjunction $Q_j^{(g, f)}(\bm X_F)$ in \eqref{eqn:prototypical} has at most the widths $\fhtw_F$ and $\subw_F$ of $\skeleton$ in the original query~\eqref{eqn:Q:R:F}.
Query~\eqref{eqn:prototypical} can be solved in time~\eqref{eqn:after:idea:1} using $\InsideOut$ or \eqref{eqn:panda:final} using $\panda$.

%% file: relatedwork.tex
\section{Related Work}
\label{sec:relatedwork}

\subparagraph{Color-coding.}
The color-coding technique~\cite{DBLP:journals/jacm/AlonYZ95} underlies existing approaches to answering queries with disequalities~\cite{Papadimitriou:JCSS:99,Bagan:CSL:07,Koutris:TCS:17}, the homomorphic embedding problem~\cite{DBLP:series/txtcs/FlumG06}, and motif finding and counting in 
computational biology~\cite{DBLP:journals/tit/AlonBNNR92}. This technique has been originally proposed for checking cliques of inequalities. It is typically used in conjunction with a dynamic programming
algorithm, whose analysis involves combinatorial arguments that
make it difficult to apply and generalize to problems beyond the path query from Eq.~\eqref{eq:intro:Q2}. 
For example, it is unclear how to use color coding to recover the Plehn and
Voigt result~\cite{PlehnV90} for the induced path query from Eq.~\eqref{eq:intro:Q3}. In this paper, we
generalize the technique to arbitrary conjunctions of $\nae$ predicates and from graph coloring to hypergraph coloring.

\subparagraph{Queries with disequalities.}
Our work also generalizes prior work on answering queries with disequalities, which are a special case of queries with negated relations of bounded degree. 

Papadimitriou and Yannakakis~\cite{Papadimitriou:JCSS:99} sho\-wed that any acy\-clic join query $Q$ with an arbitrary set of disequalities on $k$ variables can be evaluated in time $2^{O(k \log k)}\cdot |D|\cdot |Q(D)|\cdot\log^2 |D|$ over any database $D$. 
This builds on, yet uses more colors than the color-coding technique. 

Bagan et al~\cite{Bagan:CSL:07} extended this result to free-connex acyclic queries; they also shaved off a $\log |D|$ factor by using a RAM model of computation differently from ours, where sorting can be done in linear time.

Koutris et al~\cite{Koutris:TCS:17} introduced a practical algorithm for conjunctive queries with disequalities: Given a select-project-join (SPJ) plan for the conjunctive query without disequalities, the disequalities can be solved uniformly using an extended projection operator. The reliance on SPJ plans is a limitation, since it is known that such plans are suboptimal for join processing~\cite{skew} and are inadequate to achieve the $\fhtw$ and $\subw$ complexity bounds. Our approach uses the $\InsideOut$~\cite{faq} and $\panda$~\cite{panda} query evaluation algorithms and inherits their low data complexity, thus achieving both bounds as stated in Theorem~\ref{thm:complexity}. 

Differently from prior work and in line with our work, Koutris et al~\cite{Koutris:TCS:17} also investigated query structures for which the combined complexity becomes polynomial: This is the case for queries whose augmented hypergraphs have bounded treewidth (an augmented hypergraph is the hypergraph of the skeleton conjunctive query extended with one hyperedge per disequality). 
Koutris et al~\cite{Koutris:TCS:17} further proposed an alternative query answering approach that uses the probabilistic construction of the original color-coding technique coupled with any query evaluation algorithm. When restricted to queries with disequalities, our query complexity analysis is more refined than~\cite{Koutris:TCS:17} as the number of colors used in our generalization of color-coding is sensitive to the query structure.

\subparagraph{Tensor decomposition.}
Our Boolean tensor decomposition for conjunctions of $\nae$ predicates draws on the general framework of tensor decomposition used in signal processing and machine learning~\cite{TensorDecomp:2009,TensorDecomp:2017}. It is a special case of sum-product decomposition and  
a powerful tool. Typical dynamic programming algorithms solve subproblems by {\em
combining} relations and {\em eliminating} variables~\cite{Yannakakis:VLDB:81,FDB:TODS:2015,faq}. 
The sum-product decomposition is the dual approach that {\em decomposes} a formula and
{\em introduces} new variables.
The $\panda$ algorithm~\cite{panda} achieves a
generalization of the submodular width by rewriting a conjunction as a
sum-product over tree decompositions. By combining $\panda$ with our
Boolean tensor decomposition, we can answer queries with negation in time defined by the submodular width.

While close in spirit to $k$-{\sf restrictions}~\cite{DBLP:journals/talg/AlonMS06}, our approach to derandomization of the construction of the Boolean tensor decomposition is different since we would like to execute it in time
defined by the $\fhtw$-bound for computing $\skeleton$. Our derandomization uses a code-concatenation technique where the outer-code is a linear error-correc\-ting code on the Gilbert-Varshamov boundary~\cite{DBLP:journals/tit/PoratR11} that can be constructed in linear time.
As a byproduct, the code enables an efficient construction of an $(N,k^2,k)$-perfect hash family of size $O(k^2\log N)$. To the best of our knowledge, the prior
constructions yield families of size $O(k^4\log N)$~\cite{DBLP:journals/talg/AlonMS06}.

\subparagraph{Data sparsity.}
We connect two notions of sparsity in this work. One is the bounded degree of the input relations that are negated in the query. There are notions of sparsity beyond bounded degree, cf.\@~\cite{Sparsity:2017} for an excellent and comprehensive course on sparsity. The most refined sparsity notion is that of {\em nowhere denseness}~\cite{Grohe:JACM:2017}, which characterizes the {\em input monotone} graph classes on which FO model checking is fixed-parameter trac\-table. We leave as future work the generalization of our work to queries with negated nowhere-dense relations.

The second notion of sparsity used in this work is given by the Boolean tensor rank of the Boolean tensor decomposition of the conjunction of $\nae$ predicates.
We note that the relation represented by such a conjunction is not necessarily nowhere dense.

%% file: conclusions.tex
\section{Concluding remarks}
\label{sec:conclusions}

In this paper, we studied the complexity of answering conjunctive queries with negation on relations of bounded degree. We give an approach that matches the data complexity of the best known query evaluation algorithms $\InsideOut$~\cite{faq} and $\panda$~\cite{panda}. 

An intriguing venue of future research is to further lower the query complexity of our approach. 
Proposition~\ref{prop:theta:lower:bound} presented lower bounds on
$\theta^*$ that are dependent on the structure of the multi-hypergraph $\calG$ of the input query.
It is an intriguing open problem to give a lower bound on $\theta^*$ 
that is dependent on some known parameter of $\calG$.
Appendix~\ref{sec:query:complexity} discusses two further ideas on how to reduce the query complexity: 
\begin{itemize}
\item Cast coloring as a join of ``coloring predicates'' and apply the $\InsideOut$ algorithm on the resulting query with the coloring predicates taken into account; and 
\item Exploit symmetry  to answer the $k$-path query in
time $2^{O(k)}N\log N$~\cite{DBLP:journals/jacm/AlonYZ95} instead of $O(k^k N\log N)$.
\end{itemize}
Our approach extends immediately to unions of conjunctive queries with negated relations and of degree bounds on the positive relations. In the latter case, we can achieve a runtime depending on the {\em degree-aware} version of the submodular width~\cite{panda}.

We finally note that our Boolean tensor decomposition technique cannot be generalized to more powerful semirings such as the sum-product semiring over the reals due to an intrinsic computational difficulty: 
The counting version of the (induced) $k$-path query from Section~\ref{sec:intro} 
is $\SWONE$-hard~\cite{DBLP:conf/icalp/ChenTW08,DBLP:series/txtcs/FlumG06}.

%% file: appendix.tex
\section{Further Preliminaries}

In this section we give background on approximating distributions and on the $\InsideOut$ and $\panda$ algorithms for answering queries.

\subsection{Approximating distributions}
\label{subsec:approx-distrib}

Let $\calD$ and $\calP$ be two distributions on the same set of variables $\bm
X$, with probability mass functions $p_\calD$ and $p_\calP$, respectively.
Then $\norm{\calD-\calP}_1 = \sum_{\bm x}|p_\calD(\bm x)-p_\calP(\bm x)|$
is the total variational distance between the two distributions (the
$\ell_1$-norm of the difference).

Let $\calD$ be a distribution of vectors $f \in \Sigma^N$, where $\Sigma$ is a
finite alphabet. We think of $\calD$ as a distribution of $N$ random variables
$\bm X = (X_1,\dots,X_N)$, where $X_i = f(i)$. 
For any non-empty subset $I\subseteq [N]$, let
$\calD_I$ denote the marginal distribution of $\calD$ on $\bm X_I = (X_i)_{i\in
I}$.  

Let $\bm P = (p_{vj})_{v \in \Sigma, j \in [N]}$ denote a 
$|\Sigma| \times N$ matrix, where $\sum_{v \in
\Sigma} p_{vj} =1$ for all $j \in [N]$. This matrix specifies a 
distribution $\calD^{\bm P}$ on 
$\Sigma^N$ called the {\em product distribution}, defined by the
following probability mass
\[ \pr[X_1=v_1,\dots,X_N=v_N] = \prod_{j=1}^N p_{v_jj}. \]
In words, $p_{vj}$ is the probability that $X_j$ takes value $v$, and the
variables are independent. 
Let $\calS \subseteq \Sigma^N$ be a multiset of vectors; then, $\calS$ is said to
be a {\em $(k,\delta)$-approximation} to $\calD^{\bm P}$ if the following holds: for
any $\ell \leq k$, any set $I \in \binom{[N]}{\ell}$, and $\bm v \in \Sigma^I$,
we have
\[ |\pr_{\calS}[\bm X_I = \bm v_I] - \pr_{\calD^{\bm P}}[\bm X_I = \bm v_I]| < \delta. \]
The probability over $\calS$ is taken over the uniform distribution on $\calS$.

\bthm[Even et al~\cite{MR1634340}]
Given the matrix $\bm P$ for a product distribution $\calD^{\bm P}$ and two parameters $k$ and $\delta$, 
a sample space $\calS \subseteq \Sigma^N$ of size
$\poly(2^k,1/\delta,\log N)$ that is a
$(k,\delta)$-approximation for $\calD^{\bm P}$ can be computed in
time $\poly(N, 2^k,1/\delta,\log |\Sigma|)$.
\label{thm:even:et:al}
\ethm

\subsection{$\faq$ and the $\InsideOut$ algorithm: Missing Details from Section~\ref{subsec:insideout-panda}}
\label{subsec:insideout}

\bdefn[Commutative semiring]
    A triple $(\bm D,\oplus,\otimes)$ is a {\em commutative semiring} if 
    $\oplus$
    and $\otimes$ are commutative binary operators over $\bm D$ satisfying the
    following:
    \begin{enumerate}
    \item $(\bm D, \oplus)$ is a commutative monoid with an additive identity, denoted
    by $\bm  0$.
    \item $(\bm D, \otimes)$ is a commutative
    monoid with a multiplicative identity, denoted by $\bm  1$.
    (In the usual semiring definition, we do not need the multiplicative monoid to be commutative.)
    \item $\otimes$ distributes over $\oplus$.
    \item For any element $e \in \bm D$, $e \otimes \bm  0 = \bm  0 \otimes e =
    \bm  0$.
    \end{enumerate}
\edefn

Recall the definition of the $\faq$ problem from Section~\ref{subsec:insideout-panda}.

Let $\partial(n)$ denotes all edges incident to $n$ in $\calH$ 
and $J_n=\cup_{S\in\partial(n)} S$. 
The idea behind variable elimination 
\cite{DBLP:journals/ai/Dechter99,MR1426261,zhangpoole94} is to
evaluate~\eqref{eqn:sum:prod} by ``folding'' common
factors, exploiting the distributive law:
\begin{footnotesize}
\begin{eqnarray*}
    \bigoplus_{x_{f+1}}\cdots
   \bigoplus_{x_n}
   \bigotimes_{S\in\calE} \psi_S(\bm x_S)
   =\bigoplus_{x_{f+1}}\cdots
      \bigoplus_{x_{n-1}}
      \bigotimes_{S\in\calE-\partial(n)} \psi_S(\bm x_S)
      \otimes \underbrace{
         \left( \bigoplus_{x_n} \bigotimes_{S\in\partial(n)}
\psi_S(\bm x_S)\right)}_{\text{new factor } \psi_{J_n-\{n\}}},
\end{eqnarray*}
\end{footnotesize}
where the equality follows from the fact that $\otimes$ distributes over $\oplus$.

The $\InsideOut$ algorithm~\cite{faq} extends variable elimination with the following observation. For any two sets $S, T\subseteq[n]$ such that $S\cap T\neq \emptyset$ and for any factor $\psi_S$,
the function 
$\psi_{S/T} : \prod_{i \in S \cap T} \dom(X_i) \to \bm D$
defined by
\[
   \psi_{S/T}(\bm x_{S \cap T}) =
   \begin{cases}
      \bm 1 & \exists \bm x_{S-T} \text{ s.t. } 
          \psi_S(\bm x_{S \cap T},\bm x_{S-T}) \neq \bm 0\\
      \bm 0 & \text{otherwise}
   \end{cases}
\]
is called the {\em indicator projection} of $\psi_S$ onto $T$. 
Using indicator factors, $\InsideOut$ computes the following factor  when marginalizing $X_n$ away:
\begin{multline}\label{eqn:sub:query}
    \psi_{J_n-\{n\}}(\bm x_{J_n-\{n\}}) = 
     \bigoplus_{x_n}\left[  \bigotimes_{S\in\partial(n)} \psi_S
     (\bm x_S) 
        \otimes \bigotimes_{\substack{S\notin \partial(n),\\S\cap J_n\neq \emptyset}}
            \psi_{S/J_n}
           (\bm x_{S\cap J_n}) 
            \right].
\end{multline}
The key advantage of indicator projections is that the intermediate factor
$\psi_{J_n-\{n\}}$ can be computed using any 
{\em worst-case optimal join} algorithms~\cite{LFTJ,NPRR12,skew,anrr} 
in time $O(m\cdot n\cdot\log N\cdot\agm(J_n))$, where $m=|\calE|$, $N$ is the input size in the listing representation, and $\agm(J_n)$ denotes the $\agm$-bound~\cite{AGM08} on the set $J_n$.

After computing the intermediate factor, the resulting problem is another instance 
of $\sumprod$ on a modified multi-hypergraph $\calH'$, 
constructed from $\calH$ by removing vertex $n$ (corresponding to variable $X_n$) along with
all edges in $\partial(n)$ (corresponding to the relations whose schemas contain the variable $X_n$), and {\em adding} a new hyperedge $J_n-\{n\}$.
Recursively, we continue this process until all variables $X_n,X_{n-1},\dots,X_{f+1}$
are eliminated.
At this point, we are ready to report the output $\varphi(\bm x_F)$ in time  $O(m\cdot n\cdot\log N\cdot\agm(F))$.
However, if we continue with eliminating the remaining variables $X_{f}, X_{f-1}, \ldots, X_1$, then we can subsequently report the output in time roughly $\outputsize$ where $\outputsize$ is the output size in the listing representation~\cite{faq}.
More formally, we analyze the runtime of $\InsideOut$ using variable orderings
and their widths.

\bdefn[Vertex ordering of a hypergraph]
A {\em vertex ordering} (also called {\em variable ordering}) of a hypergraph 
$\calH=(\calV,\calE)$ is a listing $\sigma = (v_1,\dots,v_n)$ of all vertices in $\calV$.
Let ${\bf\Sigma}(\calH)$ denote the set of all vertex orderings $\sigma = (v_1,\dots,v_n)$ of $\calH$,
and let ${\bf\Sigma}_F(\calH)$ denote the set of all vertex orderings $\sigma = (v_1,\ldots,v_n)$ of $\calH$
that satisfy $\{v_1,\ldots,v_f\}=F$, i.e., where free variables $F=[f]$ appear first in $\sigma$.
\label{defn:Sigma_F(H)}
\edefn

\bdefn[Elimination hypergraph sequence]\ 
Given a vertex ordering $\sigma = (v_1,\dots,v_n)$ of $\calH$, for
$j = n,n-1,\dots,1$ we recursively define a sequence of $n$ hypergraphs 
$\calH^\sigma_n,\calH^\sigma_{n-1},\dots,\calH^\sigma_1$ as follows.
Define $\calH^\sigma_n = (\calV^\sigma_n,\calE^\sigma_n) = (\calV,\calE) = \calH$ and 
\begin{eqnarray}
   \partial^\sigma(v_n) &=& \bigl\{ S \in \calE^\sigma_n \suchthat v_n \in S
   \bigr\},\label{eqn:partial:vn}\\
     J^\sigma_n &=& \bigcup_{S \in \partial^\sigma(v_n)} S. \label{eqn:Un}
\end{eqnarray}
For each $j=n-1, n-2, \dots, 1$, define the hypergraph
$\calH^\sigma_j = (\calV^\sigma_j, \calE^\sigma_j)$ as follows.
\begin{eqnarray}
    \calV^\sigma_j &=& \left\{v_1,\dots,v_j\right\} \nonumber\\
    \calE^\sigma_j &=& \left(\calE^\sigma_{j+1} - \partial^\sigma(v_{j+1})\right)
    \cup \bigl\{ J^\sigma_{j+1}  - \{v_{j+1}\} \bigr\}\nonumber\\
 \partial^\sigma(v_j)&=&\left\{ S \in \calE^\sigma_j \suchthat v_j \in S\right\}\nonumber\\
     J^\sigma_j &=& \bigcup_{S\in\partial^\sigma(v_j)} S.\label{eqn:Uj}
\end{eqnarray}
The above sequence of hypergraphs is called the {\em elimination hypergraph sequence} 
associated with the vertex ordering $\sigma$.
\label{defn:ehs}
\edefn

\bdefn[Induced $g$-width~\cite{adler:dissertation, panda}]\label{defn:inducedFECWidth}
Let $\calH=(\calV,\calE)$ be a hypergraph.
Let $g: 2^{\calV} \to \mathbb R^+$ be a function and $\sigma=(v_1,\dots,v_n)$ be a 
vertex ordering
of $\calH$. Then, the {\em induced $g$-width} of $\sigma$ is the quantity
$\max_{j\in [n]} g(J^\sigma_j).$
When $g(B) = |B|-1$, this is called the {\em induced tree-width} of $\sigma$, denoted
by $\itw(\calH,\sigma)$
When $g(B) = \rho^*_\calH(B)$ (where $\rho^*_\calH(B)$ denotes the fractional edge cover number~\cite{DBLP:journals/talg/GroheM14} of the hypergraph $\calH[B]$ which is the restriction of $\calH$ to $B$), this is called the {\em induced fractional
hypertreee width} of $\sigma$, denoted by $\ifhtw(\calH,\sigma)$.
\edefn

From the above along with the fact that $\agm(B) \leq N^{\rho^*_\calH(B)}$ for any $B\subseteq\calV$, we have

\bthm[\cite{faq}]
Given an $\faq$ query $\varphi$  with hypergraph $\calH=(\calV, \calE)$ and a variable ordering $\sigma\in{\bf\Sigma}_F(\calH)$,
the $\InsideOut$ algorithm computes the output of $\varphi$ in time $O(m\cdot n^2\cdot\log N\cdot (N^{\ifhtw(\calH,\sigma)}+\outputsize))$.
\label{thm:insideout-runtime:app}
\ethm
(Recall that $n = |\calV|$, $m=|\calE|$, and $N$ is the input size in the listing representation.)

Vertex orderings of a hypergraph $\calH$ are an alternative way to characterize {\em tree decompositions} of $\calH$. Recall the definitions of tree decomposition and {\em $F$-connex} tree decomposition from Section~\ref{subsec:insideout-panda}.

\bdefn[$g$-width and $g$-width$_F$ of a hypergraph~\cite{adler:dissertation}]
Given a hypergraph $\calH=(\calV,\calE)$ and a function $g : 2^\calV \to \mathbb R^+$,
the {\em $g$-width} of a tree decomposition $(T, \chi)$ of $\calH$ is 
$\max_{t\in V(T)} g(\chi(t))$.
The {\em $g$-width of $\calH$} is the {\em minimum} $g$-width
over all tree decompositions $(T, \chi)\in\td(\calH)$.

Given additionally a set $F\subseteq \calV$, the $g$-width$_F$ of $\calH$ is the minimum
$g$-width over all {\em $F$-connex} tree decompositions $(T,\chi)\in\td_F(\calH)$.
(Note that when $F=\emptyset$, $g$-width$_F$ becomes identical to $g$-width.)
\label{defn:g-width}
\edefn

\bdefn[Common width parameters]
Given a hypergraph $\calH$,
the {\em tree-width} of $\calH$, denoted by
$\tw(\calH)$, is the $s$-width of $\calH$ where the function $s$ is defined as $s(B)=|B|-1$.
The {\em fractional hypertree width} of $\calH$,
denoted by $\fhtw(\calH)$, is the $\rho^*_{\calH}$-width of $\calH$.

Given additionally a set $F\subseteq \calV$, we use $\tw_F(\calH)$ and $\fhtw_F(\calH)$
to denote the $s$-width$_F$ and $\rho^*_{\calH}$-width$_F$ of $\calH$ respectively.
(When $F=\emptyset$, $\tw_F(\calH)=\tw(\calH)$ and $\fhtw_F(\calH)=\fhtw(\calH)$.)

(Section~\ref{subsec:insideout-panda} provides alternative yet equivalent definitions of $\fhtw(\calH)$ and $\fhtw_F(\calH)$. See~\cite{panda} for a proof of equivalence.)

We will use an $\faq$-query $\varphi$ and its hypergraph $\calH$ interchangeably, and use 
$\tw(\varphi)$, $\fhtw(\varphi)$, $\tw_F(\varphi)$, $\fhtw_F(\varphi)$ to denote
$\tw(\calH)$, $\fhtw(\calH)$, $\tw_F(\calH)$, $\fhtw_F(\calH)$ respectively.
\label{defn:common-widths}
\edefn

A function $g: 2^{\calV} \to \mathbb R^+$ is said to be {\em monotone}
if $g(A) \leq g(B)$ whenever $A\subseteq B$.

\blmm[\cite{faq}]
Let $g: 2^\calV \to \mathbb R^+$ be a monotone function and $\calH = (\calV, \calE)$ be a hypergraph.
Then, there exists a tree decomposition of $\calH$ with $g$-width $w$ if and only if
there exists a vertex ordering $\sigma\in{\bf\Sigma}(\calH)$ such that the induced $g$-width of $\sigma$ is $w$.

Moreover given additionally a set $F\subseteq \calV$, there exists an {\em $F$-connex} tree decomposition of $\calH$ with $g$-width $w$ if and only if
there exists a vertex ordering $\sigma\in{\bf\Sigma}_F(\calH)$ such that the induced $g$-width of $\sigma$ is $w$.
\label{lmm:g-width}
\elmm
Because the functions $s(B) = |B|-1$ and
$\rho^*_\calH(B)$ are monotone, the following holds.

\bcor
Given any hypergraph $\calH = (\calV, \calE)$, we have $\tw(\calH) = \min_{\sigma\in{\bf\Sigma}(\calH)}
\itw(\calH,\sigma)$ and $\fhtw(\calH) = \min_{\sigma\in{\bf\Sigma}(\calH)}\ifhtw(\calH,\sigma)$.
Given additionally a set $F\subseteq \calV$, we have $\tw_F(\calH) = \min_{\sigma\in{\bf\Sigma}_F(\calH)}
\itw(\calH,\sigma)$ and $\fhtw_F(\calH)$ $= \min_{\sigma\in{\bf\Sigma}_F(\calH)}\ifhtw(\calH,\sigma)$.
\label{prop:vo-fhtw}
\ecor

Combining the above corollary with Theorem~\ref{thm:insideout-runtime:app},
we get Theorem~\ref{thm:insideout-runtime}.

\subsection{Submodular width and the $\panda$ algorithm: Missing Details from Section~\ref{subsec:insideout-panda}}
\label{subsec:panda}

Classic width parameters, such as the ones given by Definitions~\ref{defn:g-width} and~\ref{defn:common-widths}, are defined by first defining the width of a tree decomposition and then choosing the tree decomposition that minimizes that width.
Hence for each query, there is a single best tree decomposition that is used to define the width and compute the query.

Recently more advanced width parameters were introduced such as the {\em adaptive width}~\cite{DBLP:journals/mst/Marx11}, {\em submodular width}~\cite{Marx:subw}, and {\em degree-aware submodular width}~\cite{panda}.
Those newer width notions are more dynamic in the sense that for the same query, multiple tree decompositions are simultaneously used to compute (different parts of) the query, thus allowing for better bounds and faster query evaluation algorithms.
The newer width notions are also intimately related to information theory~\cite{panda}.

Recall the definitions of {\em polymatroids} and submodular width ($\subw$ and $\subw_F$) from Section~\ref{subsec:insideout-panda}.
Comparing these definitions to the analogous definitions of $\fhtw$ and $\fhtw_F$ in the same section, it is easy to see that:

\bcor[\cite{Marx:subw,panda}]
For any hypergraph $\calH=(\calV,\calE)$ and set $F\subseteq\calV$, we have $\subw_F(\calH)\leq \fhtw_F(\calH)$.
Moreover, there are classes of hypergraphs $\calH$ for which $\subw(\calH)\ll\fhtw(\calH)$.
\ecor

Marx~\cite{Marx:subw} proposed an algorithm to solve any Boolean conjunctive query $Q$ (i.e., when the set of free variables $F=\emptyset$) in  time $O(\poly(N^{\subw(\calH)}))$ in data complexity, where $N$ is the input data size.
The more recent $\panda$ algorithm~\cite{panda} achieves a better runtime.
See Theorem~\ref{thm:panda-runtime}.

\paragraph*{Variable Introduction and the Sum-product Decomposition}While variable elimination is the key ingredient of the $\InsideOut$ algorithm (as explained in Section~\ref{subsec:insideout}), the $\panda$ algorithm uses an additional key ingredient that complements the power of variable elimination, which is {\em variable introduction}.
In particular, consider a conjunctive query $Q(\bm X_F) \leftarrow \bigwedge_{S \in \calE} R_S(\bm X_S)$ and the corresponding $\faq$ query~\eqref{eqn:sum:prod}.
In the $\panda$ algorithm, we often decompose some relation $R_T(\bm X_T)$ into a  union of a small number of pairwise-disjoint relations $R^{(1)}_T,\ldots,R^{(k)}_T$ where typically $k=O(\log |R_T|)$.
This decomposition corresponds to {\em decomposing} the function $\psi_T(\bm X_T)$ into a {\em sum} of $k$ functions $\psi^{(1)}_T(\bm X_T),\ldots,\psi^{(k)}_T(\bm X_T)$, i.e.,
\begin{equation}
\psi_T(\bm x_T) = \bigoplus_{i\in[k]} \psi^{(i)}_T(\bm x_T).\label{eqn:decompose:R_T}
\end{equation}
Alternatively, we can {\em introduce} an additional variable $X_{n+1}$ whose domain $\dom(X_{n+1})=[k]$ and define $\overline T=T\cup\{n+1\}$ along with a new function: $\psi_{\overline T}$:
\begin{align*}
\psi_{\overline T}:\prod_{i\in \overline T}\dom(X_i)\rightarrow \bm D\\
\psi_{\overline T}(\bm x_{\overline T})= \psi^{(x_{n+1})}_{T}(\bm x_T)
\end{align*}
Now we can write
\[\psi_T(\bm x_T) = \bigoplus_{x_{n+1}\in\dom(X_{n+1})} \psi_{\overline T}(\bm x_{\overline T})\]

Correspondingly, the entire $\faq$ query $\varphi(\bm X_F)$ can be expressed as a sum of $k$ sub-queries $\varphi^{(1)}(\bm X_F), \ldots, \varphi^{(k)}(\bm X_F)$:
\begin{align*}
   \varphi(\bm x_F)
   &= \bigoplus_{x_{f+1}} \cdots \bigoplus_{x_n} \bigotimes_{S\in\calE}\psi_S(\bm x_S)\\
   &= \bigoplus_{x_{f+1}} \cdots \bigoplus_{x_n}
   \left[
   \psi_T(\bm x_T)\otimes\bigotimes_{S\in\calE-\{T\}}\psi_S(\bm x_S)
   \right]\\
   &= \bigoplus_{x_{f+1}} \cdots \bigoplus_{x_n}
   \left[
   \left(\bigoplus_{x_{n+1}} \psi_{\overline T}(\bm x_{\overline T})\right)
   \otimes\bigotimes_{S\in\calE-\{T\}}\psi_S(\bm x_S)
   \right]\\
   &= \bigoplus_{x_{f+1}} \cdots \bigoplus_{x_n}\bigoplus_{x_{n+1}}
   \left[
   \psi_{\overline T}(\bm x_{\overline T})
   \otimes\bigotimes_{S\in\calE-\{T\}}\psi_S(\bm x_S)
   \right]\\
   &= \bigoplus_{x_{n+1}\in[k]}\;
   \underbrace{\bigoplus_{x_{f+1}} \cdots \bigoplus_{x_n}
   \bigotimes_{S\in\calE-\{T\}\bigcup\left\{\overline T\right\}}\psi_S(\bm x_S)}_{\varphi^{\left(x_{n+1}\right)}(\bm x_F)}
\end{align*}
Now, we have an $\faq$ query over a larger hypergraph $\overline\calH=(\overline\calV=[n+1],\overline\calE)$ where $\overline\calE=\calE-\{T\}\bigcup\left\{\overline T\right\}$.
However if the decomposition~\eqref{eqn:decompose:R_T} was applied carefully,
then each one of the queries $\varphi^{(1)}(\bm X_F), \ldots, \varphi^{(k)}(\bm X_F)$ can be solved more efficiently than the original query $\varphi(\bm X_F)$.
The $\panda$ algorithm can be described as a sequence of interleaving variable eliminations and introductions.
The sequence is carefully designed in light of entropic bounds of the query. See~\cite{panda} for more details.

\paragraph*{Effect of query rewriting on width parameters}
In this section, we show that the query rewrites we apply in  Section~\ref{sec:results} do not increase the fractional hypertree width and submodular width of the input query.

We start with a simple proposition.
\bprop
For any hypergraph $\calH=(\calV=[n], \calE)$ and any $F\subseteq\calV$, we have $\subw_F(\calH)\geq 1$.
\label{prop:subw>=1}
\eprop
\bp
Consider the function $\tilde h:2^{[n]}\rightarrow R^+$ defined as follows.
For any $B\subseteq [n]$,
\[
\tilde h(B) =
\begin{cases}
1 \quad\text{ if $1 \in B$},\\
0 \quad\text{ otherwise}.
\end{cases}
\]
It is straightforward to verify that $\tilde h \in \Gamma_n\cap \ed_\calH$. Hence
\[\subw_F(\calH) \geq \quad\min_{(T,\chi)\in\td_F(\calH)}\quad\max_{t\in V(T)}\tilde h(\chi(t)) \geq 1.\]
\ep

The following lemma is used in the proof of Proposition~\ref{prop:reduced:form}.
In particular, the lemma proves that $\skeleton_i$ in \eqref{eqn:reduced:form:detailed} and \eqref{eqn:reduced:form} has at most the same fractional hypertree width and submodular width of $\skeleton$.
While this is obvious for the fractional hypertree width, it is less obvious for the submodular width.
\blmm
Given a positive integer $n$, a hypergraph $\calH=(\calV=[n],\calE)$ and a set $F\subseteq\calV$, let $n'>n$ be another integer and $\calH'=(\calV'=[n'], \calE')$ be a hypergraph satisfying the following: $\calE'= \calE \cup\left\{S_j \suchthat j\in[n']-[n]\right\}$ where for each $j\in[n']-[n]$, $S_j=\{i, j\}$ for some $i\in[n]$.
Then, we have
$\fhtw_F(\calH')\leq \fhtw_F(\calH)$ and $\subw_F(\calH')\leq\subw_F(\calH)$.
\label{lmm:subw_i<=subw}
\elmm
\bp
Given any $F$-connex tree decomposition $(T,\chi)$ of $\calH$, let $(\overline T,\overline \chi)$ be a tree decomposition of $\calH'$ that is constructed from $(T,\chi)$ as follows:
For each edge $S_j=\{i,j\}\in\calE'-\calE$ where $j\in[n']-[n]$ (hence $i\in[n]$),
we create a new bag $t_j$ with $\chi(t_j)=\{i,j\}$ and connect $t_j$ to an arbitrary bag $t\in V(T)$ where $i\in\chi(t)$.
It is easy to verify that at the end we get a valid $F$-connex tree decomposition of $\calH'$.
We denote this tree decomposition by $(\overline T,\overline \chi)$.

Let $(T^*,\chi^*)$ be an $F$-connex tree decomposition of $\calH$ whose fractional hypertree width is the minimum, i.e., $\fhtw_F((T^*,\chi^*))=\fhtw_F(\calH)$.
It follows that $(\overline T^*,\overline \chi^*)$ is an $F$-connex tree decomposition of $\calH'$
and $\fhtw_F(\calH')\leq \fhtw_F((\overline T^*,\overline \chi^*)) \leq \fhtw_F((T^*,\chi^*))=\fhtw_F(\calH)$.

Let $\beta:(\Gamma_n\cap\ed_\calH)\rightarrow \td_F(\calH)$ be a function that maps any given $h\in \Gamma_n\cap\ed_\calH$ to an $F$-connex tree decomposition $\beta(h)=(T_h,\chi_h)\in\td_F(\calH)$ that satisfies
\[\max_{t\in V(T_h)}h(\chi_h(t)) =\min_{(T,\chi)\in\td_F(\calH)}\quad\max_{t\in V(T)}h(\chi(t)).\]
Let $\beta':(\Gamma_{n'}\cap\ed_{\calH'})\rightarrow \td_F(\calH')$ be a function that maps any given $h'\in \Gamma_{n'}\cap\ed_{\calH'}$ to an $F$-connex tree decomposition $\beta'(h')=(T_{h'},\chi_{h'})\in\td_F(\calH')$ defined as follows. Let $h:2^{[n]}\rightarrow\R^+$ be the restriction of $h'$ to $2^{[n]}$. Clearly, $h\in\Gamma_n\cap\ed_\calH$. Let $\beta(h)=(T_h,\chi_h)$, then $\beta'(\calH')=(T_{h'},\chi_{h'})=(\overline T_h,\overline\chi_h)$.

Now we have
\begin{align*}
\subw_F(\calH') 
&= \max_{h'\in\Gamma_{n'}\cap \ed_{\calH'}}\quad\min_{(T,\chi)\in\td_F(\calH')}\quad\max_{t\in V(T)}h'(\chi(t))\\
&\leq \max_{h'\in\Gamma_{n'}\cap \ed_{\calH'}}\quad \max_{t\in V(T_{h'})}h'(\chi_{h'}(t))\\
&\leq \max_{h'\in\Gamma_{n'}\cap \ed_{\calH'}}\quad \max\left(1, \max_{\substack{t\in V(T_{h'}),\\\chi_{h'}(t)\subseteq[n]}}h'(\chi_{h'}(t))\right)\\
&= \max_{h\in\Gamma_{n}\cap \ed_{\calH}}\quad\max\left(1, \max_{t\in V(T_h)}h(\chi_h(t))\right)\\
&= \max\left(1, \max_{h\in\Gamma_{n}\cap \ed_{\calH}}\quad\max_{t\in V(T_h)}h(\chi_h(t))\right)\\
&= \max(1, \subw_F(\calH))\\
&\leq \subw_F(\calH)
\end{align*}
The last inequality above follows from Proposition~\ref{prop:subw>=1}.
\ep

\subsection{$k$-restriction and error-correcting codes}
\label{subsec:prelim:k-restrict}

{\em $k$-restriction} is a very general class of problems introduced by Alon,
Moshkovitz, and Safra~\cite{DBLP:journals/talg/AlonMS06}. 
An instance of a $k$-restriction problem has as
its input an alphabet $\Sigma$, a positive integer $N$ called the ``length'',
and a set $D$ of functions $d : \Sigma^k \to \{0,1\}$.
Functions in $D$ are called ``demands,'' satisfying the condition that
for every $d \in D$, there is some tuple $\bm a \in \Sigma^k$ for which $d(\bm
a)=1$, i.e., every demand is satisfiable.

The problem is to construct a family $\calF \subseteq \Sigma^N$ as small as
possible such that, for every subset $S \in \binom{[N]}{k}$ and every demand
$d \in D$, there is some $f \in \calF$ for which $d(f(S))=1$. Here, $f(S)$
denote the restriction of $f$ onto coordinates in $S$. In particular, $f(S) \in
\Sigma^k$.

It is also not hard to see that constructing a $k$-disjunct matrix is a
$(k+1)$-restriction problem, where the demands are of the form
$d(\bm a)=1$ for every $(k+1)$-tuple $\bm a$ with one $1$ and $k$ $0$s.

Let $n$ and $\Delta$ be positive integers, and $\Sigma$ a finite alphabet. 
A {\em code} of length $n$ and distance
$\Delta$ is a set of vectors $C \subseteq \Sigma^n$ such that the Hamming distance
between every two different vectors in $C$ is at least $\Delta$. The vectors
$\bm c = (c_i)_{i\in [n]}$ in $C$ are called {\em codewords}; and the indices
$i \in [n]$ are often called the {\em positions} of the code. 
The quantity $\delta = \Delta/n$ is called the {\em relative distance} of the
code.

A code $C$ is called an
{\em $[n,d,\Delta]_q$-linear code} if $C$ is a linear subspace of $\F_q^n$ of
dimension $d$ and minimum Hamming distance $\Delta$ (where $\F_q$ is the finite field of order $q$). The alphabet is
$\Sigma=\F_q$. A linear code $C$ can be generated from a {\em generator matrix} $\bm
A \in \F_q^{d \times n}$ in the sense that $C = \{ \bm A \bm m \suchthat \bm m
\in \F_q^d\}$.
(The canonical example of linear codes is the {\em Reed-Solomon
code}~\cite{MR0127464}, widely used both in theory and practice.)
One important aspect of linear codes is that, given the generator matrix $\bm A$
(which can be stored with $nd\log q$ bits), and a ``message'' $\bm m \in
\F_q^d$, we can compute any position of the codeword $\bm A \bm m$ in time
$O(d\log q)$.

\section{Missing Proofs}

\subsection{Theorem~\ref{thm:prob:const}}
\label{proof:thm:prob:const}

To prove the theorem, we need an auxiliary lemma.

\blmm
Given positive integers $2 \leq k < N$, there exists a linear $[n,d,\delta n]_q$-code 
with relative distance $\delta \geq 1-2/k^2$, alphabet size $q = \Theta(k^2)$,
length $n=O(k^2\log N)$, and dimension $d = \lceil \log_q N \rceil$. Furthermore, a
generator matrix for the code can be constructed in $O(nq^d) = O(nN)$ time.
\label{lmm:pr:code}
\elmm
\bp
For $q \geq 2$ define the $q$-ary entropy function
\[
   H_q(x) = x \log_q(q-1)-x\log_qx-(1-x)\log_q(1-x).
\]
Using the method of conditional expectation to construct a generator matrix for
linear codes, Porat and Rothschild~\cite{DBLP:journals/tit/PoratR11} were able to
obtain the following result.
Theorem 3 from~\cite{DBLP:journals/tit/PoratR11} states that, given prime power 
$q$, relative distance $\delta \in (0,1)$, integers $d,n$ such that
$d \leq (1-H_q(\delta))n$, then we can construct the generator matrix
of an $[n,d,\delta n]_q$-linear code in time $O(nq^d)$.
To apply their result in our setting, we set $q$ to be a power of $2$ in the
interval $[2k^2,4k^2)$, $d = \lceil \log_q N \rceil$, and $\delta = 1-2/k^2$.
Then, there exists the desired code with length $n = \left\lceil
\frac{d}{1-H_q(\delta)} \right\rceil$. We prove that $n = O(k^2\log N)$.
To see this, observe that
\begin{align*}
   1-H_q(\delta) 
   &= \frac{\log q
   -\delta\log(q-1)+\delta\log\delta+(1-\delta)\log(1-\delta)}{\log q}\\
   &= \frac{k^2\log\frac{q}{q-1}
   +2\log \frac{2(q-1)}{k^2}-(k^2-2)\log \frac{k^2}{k^2-2}}{k^2\log q}\\
   &= \frac{\Theta(1)}{k^2\log k}
\end{align*}
In the above, we used the fact that, for any constant $c>0$, the function
$(1+c/x)^x$ is sandwiched between $1+c$ and $e^c$. For example, since $k^2 \leq q-1 \leq
4k^2$, we have
\begin{align*}
   k^2\log \frac{q}{q-1}
   &= \log \left( 1+\frac{1}{q-1} \right)^{k^2}
   \leq \log \left( 1+\frac{1}{k^2} \right)^{k^2} \leq 1,\\
   k^2\log \frac{q}{q-1}
   &= \log \left( 1+\frac{1}{q-1} \right)^{k^2}
   \geq \log \left( 1+\frac{1}{4k^2} \right)^{k^2} \geq \log (5/4).
\end{align*}
The other two terms $2\log \frac{2(q-1)}{k^2}$ and 
$(k^2-2)\log \frac{k^2}{k^2-2}$ are bounded similarly.
\ep

The derandomization step in the proof of the theorem below is a simple adaptation of
the construction from~\cite{DBLP:journals/talg/AlonMS06}.
Their construction does not work directly on our problem because ours is not
exactly a $k$-restrictions problem, and we had to choose the ``outer code''
carefully to have good parameters with a linear runtime.

\paragraph*{Theorem~\ref{thm:prob:const} re-stated}
    Let $\calG = (U, \calA)$ be a multi-hypergraph,
    $c = \max_{h \in \col(\calG,N)}\chi(h(\calG))$, and
    $\bm p =(p_1,\dots,p_c) \in \R_+^c$ be a fixed
    non-negative real vector such that $\norm{\bm p}_1=1$.
    Define
    \begin{equation}
    \theta(\bm p) = \min_{h \in \col(\calG,N)} 
    \sum_{\bm \mu\in T_c(h)} n(\bm \mu,h) \prod_{i=1}^c p_i^{\mu_i}
    \label{eqn:theta:p:restated}
    \end{equation}
    Then, the following hold:
    
    \bi
    \item[(a)] There exists a family $\calF$ of functions $f : [N] \to [c]$ satisfying
    \eqref{eqn:good:condition} such that
    \begin{equation}
    |\calF| \leq \left\lceil \frac{\ln P(\calG,N)}{\theta(\bm p)} \right\rceil
    \leq
    \frac{|U| \log N}{\theta(\bm p)}.
    \label{eqn:F:upper:bound:restated}
    \end{equation}
    \label{thm:chromatic:restated}
    \item[(b)] There is a strongly explicit family $\calF'$ of functions $f : [N] \to [c]$ 
    satisfying~\eqref{eqn:good:condition} such that
    \begin{equation}
    |\calF'| = O\left( \frac{|U|^3 \cdot \log |U| \cdot \log N}
    {\theta(\bm p)}
    \right).
    \end{equation}
    \ei
\bp[Proof of Theorem~\ref{thm:prob:const}]
   To prove part $(a)$,
   let $h : U \to [N]$ be an arbitrary $N$-coloring of $\calG$. 
   Let $\calD$ be the distribution of functions $f : [N]
   \to [c]$ in which we assign $f(x) = i\in [c]$ with probability $p_i$ 
   independently for every $x \in [N]$. 
   In other words, $\calD$ is the product distribution specified by the matrix
   $\bm P = (p_{ij})_{i\in [c],j\in [N]}$ where $p_{ij} = p_i$ for all $j$.
   (See Section~\ref{subsec:approx-distrib}.)
   It is straightforward to verify that 
\begin{align*} 
   & \pr_{f}\left[ f \circ h \text{ is a proper coloring of } \calG \right] 
   = \pr_{f}\left[ f \text{ is a proper coloring of } h(\calG) \right] \\
   &= \sum_{\bm \mu\in T_c(h)} n(\bm \mu,h) \prod_{i=1}^c p_i^{\mu_i}
   \geq \theta(\bm p).
\end{align*}
   Now, let $\calF$ be a (random) family of functions $f : [N] \to [c]$ 
   constructed by picking randomly  and independently (with replacement) 
   $r=\left\lceil\frac{\ln P(\calG, N)}{\theta(\bm p)}\right\rceil$ functions $f$ from the distribution 
   $\calD$ above. Then, for a fixed proper $N$-coloring $h$ of $\calG$,
\begin{multline}
   \pr_{\calF}\left[\forall f \in \calF : f\circ h \text{ is not a proper coloring of } \calG 
   \right]
   \leq (1-\theta(\bm p))^r < e^{-\theta(\bm p) r}.
\end{multline}
The last inequality above holds because $1-x < e^{-x}$ for all $x\neq 0$. By the union bound, it follows that
\begin{multline}
   \pr_{\calF}\bigl[\exists h \forall f \in \calF: f\circ h \text{ is not a proper coloring of } \calG\bigr] 
   < P(\calG,N) \cdot e^{-\theta(\bm p) r} \leq 1.
\end{multline}
   Hence, there exists a family $\calF$ satisfying~\eqref{eqn:good:condition}
   whose size is bounded in~\eqref{eqn:F:upper:bound:restated}, proving part $(a)$ of
   the theorem.
   We also used the trivial fact that $P(\calG,N) \leq N^{|U|}$ to obtain the
   second inequality in~\eqref{eqn:F:upper:bound:restated}, but this inequality may 
   be a huge overestimate for some graphs $\calG$.

   To show part $(b)$, we first prove a claim.

{\bf Claim.} In $O(N^{|U|})$-time, we can construct a family $\calF$
   satisfying~\eqref{eqn:good:condition} with size 
   $|\calF| \leq 2\frac{|U| \log N}{\theta(\bm p)}$.

   {\em Proof of the claim.}
   We show how the above randomized construction of 
   $\calF$ can be derandomized to run in time $O(N^{|U|})$.
   First, we show how to construct the family $\calF$ 
   satisfying~\eqref{eqn:F:upper:bound:restated} in an {\bf un}reasonable amount of time.
   To do so, we formulate the construction of $\calF$ as a {\sf set cover} problem.
   In this set cover problem, the universe consists of all $N$-colorings $h : U \to
   \calG$, and every function $f : [N] \to [c]$ is a ``set,'' which contains all
   elements $h$ in the universe for which $f \circ h$ is a proper $c$-coloring
   of $\calG$. By running the greedy set-covering algorithm for this {\sf set
   cover} instance\footnote{At each step, select the set covering the most
   number of uncovered elements in the universe.}, we can construct $\calF$
   satisfying~\eqref{eqn:good:condition} in time $O(N^{|U|} \cdot c^N)$.
   From Proposition 3 from~\cite{DBLP:journals/talg/AlonMS06}, the size of
   the cover, $|\calF|$, can also be bounded by~\eqref{eqn:F:upper:bound:restated}. 
   The runtime of $O(N^{|U|}\cdot c^N)$ is too large, however.
   Of the two factors, $c^N$ is the much more serious one, caused by the fact that
   there are too many ``sets'' to choose from. The next idea is to use
   distribution approximation to reduce the number of ``sets''.

   From Theorem~\ref{thm:even:et:al}, by setting
   $\delta=\frac{\theta(\bm p)}{2c^{|U|}}$, we can construct a set $\calS
   \subseteq \Sigma^N$ with $|\calS| = \poly(c^{|U|}/\theta(\bm p), \log N)$ 
   for which $\calS$ is 
   a $(|U|,\frac{\theta(\bm p)}{2c^{|U|}})$-approximation of $\calD$. 
   The algorithm runs in time $\poly(N,c^{|U|}/\theta(\bm p))$.
   Let $I = \{ h(j) \suchthat j \in U\}$. 
   Recall that a random $f \in \Sigma^N$ (from either $\calD$ or $\calS$)
   is equivalent to a random tuple $\bm X_{[N]}$ where $X_i = f(i)$.
   Let $\bm V_I \subseteq \Sigma^I$
   denote the set of tuples $\bm v_I \in \Sigma^I$ such that $f \circ h$ is a
   proper $c$-coloring of $\calG$ iff $\bm X_I = \bm v_I$.

   Due to the $(|U|, \frac{\theta(\bm p)}{2c^{|U|}})$-approximation, it follows that
   \begin{align*}
      &\pr_{f \sim \calS}[f \circ h \text{ is a proper coloring of } \calG]
      = \sum_{\bm v_I \in \bm V_I} \pr_{f \sim \calS}[\bm X_I = \bm v_I]\\
      &\geq \sum_{\bm v_I \in \bm V_I} \left( \pr_{f \sim \calD}[\bm X_I = \bm v_I] 
      - \frac{\theta(\bm p)}{2c^{|U|}} \right)\\
      &=\pr_{f \sim \calD}[f \circ h \text{ is a proper coloring of } \calG]
      - |\bm V_I| \frac{\theta(\bm p)}{2c^{|U|}}
      \geq \theta(\bm p) - c^{|I|} \frac{\theta(\bm p)}{2c^{|U|}}
      \geq \theta(\bm p) - c^{|U|} \frac{\theta(\bm p)}{2c^{|U|}}\\
      &=\theta(\bm p)/2.
   \end{align*}
   Now, we run the greedy algorithm on the smaller set collection $\calS$, and 
   apply Proposition 3 from~\cite{DBLP:journals/talg/AlonMS06}, we
   obtain a family $\calF$ whose size is bounded by $\frac{2 |U|\log
   N}{\theta(\bm p)}$. 
   The greedy algorithm now runs in time $\poly(N^{|U|}, c^{|U|}/\theta(\bm p))$ 
   only, because the ``sets'' are now only members of $|\calS|$ instead of $[c]^N$.
   This is an improvement that we will use as a blackbox for part $(b)$.
   {\em This concludes the proof of the above claim.}
   \qed
   
   To prove part $(b)$ of Theorem~\ref{thm:prob:const}, our next idea is to use ``code concatenation'' to 
   reduce the runtime of 
   $\poly(N^{|U|}, c^{|U|}/\theta(\bm p))$ 
   dramatically down to $\tilde O(N)$.
   Code concatenation is an idea widely used in coding theory and
  derandomization~\cite{DBLP:journals/talg/AlonMS06,DBLP:conf/focs/NaorSS95}, and their
  numerous applications such as group testing~\cite{DBLP:conf/icalp/NgoPR11,DBLP:conf/soda/IndykNR10} or
   compressed sensing~\cite{DBLP:conf/stacs/NgoPR12}.

   Let $C$ denote the $[n,d,\delta n]_q$-linear code obtained from 
   Lemma~\ref{lmm:pr:code} with $k=|U|$ and $n=O(k^2\log N)$.
Now, every pair of codewords share at most $(1-\delta)n = 2n/|U|^2$ positions. 
Hence, given any
collection of $|U|$ codewords, there has to be $n -
\binom{|U|}{2}\frac{2n}{|U|^2}>0$
positions where those $|U|$ codewords contain 
pair-wise distinct symbols. In particular, 
there has to be one position $i \in [n]$ in which all $|U|$ codewords have different symbols.
%

From the derandomization result in the claim above, applied to the same problem
with $N$ replaced by $q$,
we know that we can construct a family
$\bar \calF$ of functions $\bar f : [q] \to [c]$ such that, for any proper
$q$-coloring $\bar h : U \to [q]$ of $\calG$, there exists an $\bar f \in
\bar \calF$ such that $\bar f \circ \bar h$ is a proper $c$-coloring of
$\calG$. The family $\bar \calF$ has size bounded by $$|\bar \calF| \leq
\frac{2|U| \log q}{\theta(\bm p)} = O\left(\frac{|U| \log |U|}{\theta(\bm p)}\right).$$
Now, we construct $\calF$ by using $C$ as the outer code and $\bar\calF$ as
the inner code. In particular, let $C = \{ \bm w_j \suchthat j \in [N]\}$.
For every position $i \in [n]$ of $C$ and every function
$\bar f \in \bar \calF$, there is a function $f : [N] \to [c]$ defined by
$f(j) = \bar f(\bm w_j(i))$.
(Here, $\bm w_j$ denotes the $j$th codeword and $\bm w_j(i)$ is the symbol at the
$i$th position of the $j$th codeword.)
We show that this family $\calF$ satisfies condition~\eqref{eqn:good:condition}.
Let $h : U \to [N]$ be an arbitrary $N$-coloring of $\calG$.
Then, as reasoned above, there has to be a position $i \in [n]$ for which
the symbols $\bm w_j(i)$ are distinct for all $j \in h(U)$.
Define $\bar h : U \to [q]$ where $\bar h(v) = \bm w_{h(v)}(i)$.
Then, $\bar h$ is a proper $q$-coloring of $\calG$, which means there exists
$\bar f \in \bar \calF$ for which $\bar f \circ \bar h$ is a proper $c$-coloring
of $\calG$. By construction $\bar f \circ \bar h \in \calF$.
The size of the strongly explicit family $\calF$ constructed above is bounded 
by $|U|^2 \log N \cdot |\bar \calF|$.
%
\ep

\subsection{Proposition~\ref{prop:theta:lower:bound}}
\label{proof:prop:theta:lower:bound}
\paragraph*{Proposition~\ref{prop:theta:lower:bound} re-stated}
    The following hold:
    \bi
    \item[(a)] Given a multi-hypergraph $\calG=(U,\calA)$ with $|U|=k$ and $c = \max_{h \in
        \col(\calG,N)}\chi(h(\calG))$, it holds that
    $\theta^* \geq \frac{1}{c^c} \geq \frac{1}{k^k}$.
    \item[(b)] Suppose $\calG = K_{\bm \mu}$ for some positive integer tuple $\bm
    \mu = (\mu_1,\dots,\mu_\ell)$, where $\mu_1\geq \cdots \geq \mu_\ell \geq 1$. 
    Let $S_\ell$ denote the set of all permutations of $[\ell]$, and 
    $\fp(\bm \mu)$ denote the number of permutations $\pi \in S_\ell$ for which
    $\mu_i = \mu_{\pi(i)},\forall i \in [\ell]$. Then,
    \begin{align}
    \theta^* = \max_{\bm p} \sum_{\pi \in S_\ell} \prod_{i=1}^\ell
    p_i^{\mu_{\pi(i)}} 
    \geq \sum_{\pi\in S_\ell} \prod_{i=1}^\ell \left(\frac{\mu_i}{\norm{\bm
            \mu}_1}\right)^{\mu_{\pi(i)}} 
    \geq \fp(\bm \mu) \prod_{i=1}^\ell \left(\frac{\mu_i}{\norm{\bm
            \mu}_1}\right)^{\mu_{i}}.\label{eqn:theta*:combined:restated}
    \end{align}
    \ei
\begin{proof}
 Part $(a)$ is trivial to verify, by picking $\bm p=(p_i)_{i\in [c]}$ where
 $p_i=1/c$.
 For part $(b)$, consider any $N$-coloring $h$ of $\calG$. Let
 $I_1,\dots,I_\ell$ denote the independent sets of $\calG$ where $|I_i|=\mu_i$.
 It is easy to verify that $h(\calG)$ is also a complete $\ell$-partite graph,
 with independent sets $I^h_i, i \in [\ell]$ such that $I^h_i \subseteq I_i$
 for all $i \in [\ell]$.
 Every $\ell$-coloring of $h(\calG)$ is a permutation $\pi \in S_\ell$ which
 assigns color $i$ to all vertices in $I^h_i$. Hence, for any $\pi$,
 \begin{align*}
    \sum_{\bm \mu' \in T_\ell(h)} n(\bm \mu,h) \prod_{i=1}^\ell p_i^{\mu'_{i}}
    = \sum_{\pi \in S_\ell} \prod_{i=1}^\ell p_i^{I^h_{\pi(i)}}
    \geq \sum_{\pi \in S_\ell} \prod_{i=1}^\ell p_i^{I_{\pi(i)}}
    = \sum_{\pi \in S_\ell} \prod_{i=1}^\ell p_i^{\mu_{\pi(i)}}
 \end{align*}
 This proves that
    $\theta^* \geq \max_{\bm p} \sum_{\pi \in S_\ell} \prod_{i=1}^\ell
    p_i^{\mu_{\pi(i)}}$. To see that equality holds, note that for 
    the function $h$ assigning $h(v) = i$, we have $I^h_i = I_i$.
    Hence, the equality in~\eqref{eqn:theta*:combined:restated} follows.
    The first inequality in~\eqref{eqn:theta*:combined:restated} follows from setting $p_i =
    (\mu_i/\norm{\bm\mu}_1)^{\mu_i}$. The last inequality is trivial.
\end{proof}

\section{Reducing query complexity}
\label{sec:query:complexity}
Theorem~\ref{thm:complexity}, proved in Section~\ref{sec:use-tensor-decomposition}, was mainly concerned with reducing the data complexity of the evaluation problem of queries of the form~\eqref{eqn:Q:R:F}.
In this section, we introduce two ideas to further reduce the combined complexity of that problem. The first
idea is to think of colorings as a join problem. The second idea is to exploit 
symmetry in the colorings to save computation time.

\paragraph*{Colorings as a join}
In the runtime expression~\eqref{eqn:after:idea:1} above, $P(\calG,c)$ can be as
large as $c^{|U|}$, even when the graph $\calG$ itself is very simple. Our next
idea aims to reduce this factor down.
We present a couple of ideas to reduce this factor down dramatically,
by encoding the colorings $g \in \col(\calG,c)$ as the result of a join as follows.  
Introduce $|U|$ new variables $(C_i)_{i \in U}$, whose domains are
$\dom(C_i) = [c]$, for all $i \in U$.
To each hyperedge $S \in \calA$, the relation $\nae(\bm
C_S)$ encodes the fact that the coloring is valid for the hyperedge $S$.
(These relations $\nae(\bm C_S)$ can be materialized, but we don't have to
because we can verify easily if a tuple belongs to the relation.)
Then, the body of the query~\eqref{eqn:prototypical} can be equivalently formulated 
as
\begin{equation}
   \bigvee_{f \in \calF} \bigwedge_{S\in\calE} R_S(\bm X_S)
\wedge \bigwedge_{i\in U}(f(X_i)=C_i)
\wedge \bigwedge_{S\in\calA}\nae(\bm C_S).
\label{eqn:before:idea:2}
\end{equation}
Now, by using a semiring over $|\calF|$-bit vectors, the
formula~\eqref{eqn:before:idea:2} can be viewed as a $\sumprod$ formula over
$|\calF|$-bit vectors, whose hypergraph $\calH' = (\calV',\calE')$ is defined as 
follows.  The vertex set is 
$\calV' = \{X_i \suchthat i \in \calV\} \cup \{C_i \suchthat i \in U\}$.
The edge set is
$\calE' = \calE \cup \{ \{X_i,C_i\} \suchthat i \in U\} \cup \{ \bm C_S
\suchthat S \in \calA\}$.
In words, $\calH'$ is a super-graph of $\calH$, where we add to $\calH$ the
edges $\{X_i,C_i\}$ for $i \in U$, and $\bm C_S$ for $S \in \calA$.

\begin{ex}
   Consider the following query:
   \begin{multline}
      Q \leftarrow \bigwedge_{i\in[5]} R_i(X_i,X_{i+1}) \wedge
            (X_4 \neq X_1) \wedge (X_4 \neq X_2) \ \wedge (X_4 \neq X_6).
   \end{multline}
   The graphs $\calH$ and $\calH'$ are shown in Fig.~\ref{fig:H:H'}.
   \label{ex:H:H'}
   \qed
\end{ex}

\input{fig-H-H}

To evaluate~\eqref{eqn:before:idea:2}, we run $\InsideOut$ on a variable ordering 
of $\calH'$. (See Section~\ref{subsec:insideout} and~\cite{faq}.) 
Let $\pi$ be a variable ordering of $\calH'$.
Given a variable $Z$ of $\calH'$ (which could be either an ``input variable'' $X_i$
or a ``color variable`` $C_j$), let $J^\pi_Z$ denote the set of
variables involved in computing the intermediate result when $\InsideOut$ 
eliminates the variable $Z$, as in Eq.~\eqref{eqn:Un} and~\eqref{eqn:Uj} of Definition~\ref{defn:ehs}.

\begin{ex}
   For the query shown in Example~\ref{ex:H:H'}, consider the following three
   different variable orders of $\calH'$ to run $\InsideOut$ on:
   \begin{align*}
      \pi_1 & = X_4X_5X_6X_3X_2X_1C_4C_6C_2C_1\\
      \pi_2 & = C_4C_6C_2C_1X_4X_5X_6X_3X_2X_1\\
      \pi_3 & = C_4X_4X_5C_6X_6X_3C_2X_2C_1X_1.
   \end{align*}
   These variable orders correspond to different tree decompositions.
   For $\pi_1$, there is an intermediate
   relation computed over $J^{\pi_1}_{C_4} = \{X_4,X_6,X_2,X_1,C_4\}$. Since
   $C_4$ is functionally determined by $X_4$, this subproblem can be solved in
   $O(N^3)$-time, because $\rho^*_{\calH}(X_4X_6X_2X_1) = 3$.
   (See Definition~\ref{defn:inducedFECWidth}.)
   For $\pi_2$, there is an intermediate relation computed over
   $J^{\pi_2}_{X_4} = \{C_4,C_6,C_2,C_1, X_4\}$, which can be solved in time
   $O((c-1)^3 \cdot N)$. This is because once we fix a binding for $X_4$, the
   color $C_4$ is also fixed at $C_4 = f(X_4)$, and there are $c-1$ choices for
   each of the other three colors.
   For $\pi_3$, the runtime is bounded by $O(cN)$, which is better
   than both $\pi_1$ and $\pi_2$. 
   \label{ex:pi1:pi2:pi3}
   \qed
\end{ex}

Assume a variable ordering $\pi$ of $\calH'$ and let $Z$ be either an input variable $X_i$
or a color variable $C_j$. Let $J^\pi_Z|_\calV$ denote the set of input
variables in $J^\pi_Z$, and $J^\pi_Z|_U$ denote the set of color variables in
$J^\pi_Z$ which are {\em not} functionally determined by input variables in
$J^\pi_Z$.
Let $\pi|_\calV$ denote the subsequence of $\pi$ containing only the input
variables.

\bprop
For a given variable ordering $\pi$ of $\calH'$, the runtime of $\InsideOut$ is
in the order of
\begin{align*}
   \alpha\cdot\max_{Z} \left(c^{|J^\pi_Z|_U|} \cdot N^{\rho^*_\calH(J^\pi_Z|_\calV)}\right)+\beta
   \geq \alpha\cdot N^{\ifhtw(\calH, \pi|_\calV)}+\beta
   \geq \alpha\cdot N^{\fhtw(\calH)}+\beta
\end{align*}
where $\alpha=|\calE'|\cdot|\calV'|^2\cdot\log N$ and $\beta = \alpha \cdot \outputsize$.
\label{prop:io-color-bound}
\eprop
\bp
The first bound is proved in the exact same way as Theorem~\ref{thm:insideout-runtime}.
The first inequality follows from the fact that $\calH'$ is a super-graph of $\calH$. The second inequality
follows from the fact that, the induced fractional hypertree width of 
of $\pi|_\calV$ w.r.t. $\calH$ is at least the fractional hypertree width of $\calH$. (See Definition~\ref{defn:inducedFECWidth})
\ep

\begin{ex}
Consider the variable order $\pi_2$ above. For each variable $Z$ in $\pi_2$, the following table shows the sets $J^{\pi_2}_Z$, $J^{\pi_2}_Z|_\calV$, $J^{\pi_2}_Z|_U$ and the time needed to compute the subproblem over $J^\pi_Z$ and eliminate $Z$:

   \begin{tabular}{l|l|l|l|l}
      $Z$ & $J^{\pi_2}_Z$ & $J^{\pi_2}_Z|_\calV$ & $J^{\pi_2}_Z|_U$ & runtime\\
      \hline
      $X_1$ & $X_2X_1C_1$      & $X_2X_1$ & $\emptyset$ & $N$\\
      $X_2$ & $X_2X_3C_1C_2$   & $X_2X_3$ & $C_1$ & $Nc$\\
      $X_3$ & $X_3X_4C_1C_2$   & $X_3X_4$ & $C_1C_2$ & $Nc^2$\\
      $X_6$ & $X_5X_6C_6$      & $X_5X_6$ & $\emptyset$ & $N$\\
      $X_5$ & $X_4X_5C_6$      & $X_4X_5$& $C_6$ & $Nc$\\
      $X_4$ & $X_4C_4C_6C_1C_2$    & $X_4$& $C_6C_1C_2$& $Nc^3$.
   \end{tabular}

   Note that for $Z=X_1$, the set $J^{\pi_2}_Z|_U$ does not contain the color variable $C_1$ but is rather empty. This is because $C_1$ is determined by the functional dependency $C_1=f(X_1)$ where $X_1$ is in $J^{\pi_2}_Z|_\calV$.
   Moreover, $\rho^*_\calH(J^{\pi_2}_Z|_\calV =\{X_1,X_2\})=1$, and hence the runtime to eliminate $Z=X_1$ is $O(N)$.
   \qed
\end{ex}

Data complexity is more important than query complexity in
database applications and we do not want to construct a variable
ordering for which the interleaving of the color variables may increase the
data complexity unnecessarily. This was the case for variable ordering $\pi_1$
in Example~\ref{ex:pi1:pi2:pi3} above: the subsequence $\pi_1|_\calV=X_4X_5X_6X_3X_2X_1$ has an optimal induced fractional hypertree width of $\ifhtw(\calH, \pi_1|_\calV) = 1$, but the interleaving of the color
variables increases the data-complexity to be $\max_Z N^{\rho^*_\calH(J^{\pi_1}_Z|_\calV)}=N^3$, hence for $\pi_1$, the first inequality in Proposition~\ref{prop:io-color-bound} is strict! 
This observation motivates the following question:

\begin{pbm}[Color amendment problem]
   Given a good variable ordering $\sigma$ of $\calH$
with low induced
fractional hypertree width $\ifhtw(\calH, \sigma)$, how do we interleave the
color variables to obtain a variable ordering $\pi$ for $\calH'$ such that 
$\rho^*_\calH(J^\pi_{X_i}|_\calV) \leq \ifhtw(\calH, \sigma)$ for all input variables
$X_i$, and that $\max_Z |J^\pi_Z|_U|$ is as small as possible?
\end{pbm}

For example, we can simply append all color variables to the beginning of
$\sigma$ to maintain 
$\rho^*_\calH(J^\pi_{X_i}|_\calV) \leq \ifhtw(\calH, \sigma)$; this was the case with
$\pi_2$ in Example~\ref{ex:pi1:pi2:pi3} above; however, the quantity 
$\max_Z |J^{\pi_2}_Z|_U| = 3 > 1 = 
\max_Z |J^{\pi_3}_Z|_U|$. Hence $\pi_2$ induces higher query complexity than
$\pi_3$ (while both $\pi_2$ and $\pi_3$ induce the same data complexity of $N$).
From the fact that computing the treewidth is $\np$-hard, it is easy to show the following.

\bprop
{\sf Color amendment} is $\np$-hard in query complexity.
\eprop

We leave open the problem of finding an approximation algorithm for {\sf color
amendment}. We present here a simple greedy algorithm which works well in some cases.
Given an input variable ordering $\sigma$ of the hypergraph $\calH$. We first
construct the corresponding tree decomposition $(T,\chi)$ for which $\sigma$ is the
GYO-elimination order of the tree decomposition. (See Section~\ref{subsec:insideout} and~\cite{faq}, for example,
for the equivalence between variable orderings and tree decompositions.)
Then, for each $i \in U$, we add the color variable $C_i$ to every bag $\chi(t)$ 
of the tree decomposition for which $X_i \in \chi(t)$. The idea is that adding $C_i$ does not increase the cost of the bag $\chi(t)$ because $X_i$ functionally determines $C_i$ for free. The running
intersection property (RIP) (property (b) of Definition~\ref{defn:td}) is still satisfied thus far; and all but the $\nae$
hyperedges of $\calH'$ are covered by the bags of the new tree decomposition.
Next, we add more color variables to bags in order to cover the $\nae(\bm C_S)$
predicates, hence satisfy property (a) of Definition~\ref{defn:td}. 

Our objective is to minimize the maximum number of non-functionally-determined color
variables over all bags. Initially, this quantity is $0$ because all color
variables are functionally determined.
Covering the predicates $\nae(\bm C_S)$ is done greedily. 
Let $\nae(\bm C_S)$ be a 
predicate not yet covered by the current tree decomposition. According to a criteria to be specified later, we choose a bag
$\chi(t)$ to add $\bm C_S$ to. Once $\bm C_S$ is added to $\chi(t)$, we deterministically enforce
 RIP as follows: for every color variable $C_i$ that is disconnected in the
tree (i.e., where there are at least two disconnected subtrees whose bags contain $C_i$), we add $C_i$ to every bag on the (unique) path connecting the two subtrees in the tree.
Finally, our criteria for choosing the bag $\chi(t)$ that we add $\bm C_S$ to is as follows:
We pick the bag $\chi(t)$ that already contains a variable in $\bm C_S$
and increases the objective value the {\em least}.
\begin{ex}
   Consider the query from Example~\ref{ex:H:H'} whose hypergraphs $\calH$ and $\calH'$ are both depicted in Figure~\ref{fig:H:H'}.
   Given the ordering $\sigma=X_1X_2X_3X_4X_5X_6$ of $\calH$, the GYO elimination procedure over $\sigma$ produces the tree decomposition in Fig.~\ref{fig:greedy:algo}(a).
   We add color variables $C_i$ to every bag containing $X_i$, resulting in the tree decomposition in Fig~\ref{fig:greedy:algo}(b).
   Note that the hyperedge $\{C_1,C_4\}$, for example, is not contained in any bag of this tree decomposition.
   To fix this, we choose the bag $\{X_1,X_2,C_1,C_2\}$ and add $\{C_1,C_4\}$ to it, resulting in the tree decomposition in Fig.~\ref{fig:greedy:algo}(c).
   Now this tree decomposition has a different problem which is that there are two disconnected subtrees of bags containing $C_4$.
   To fix this, we add $C_4$ to the middle bag $\{X_2,X_3,C_2\}$, resulting in the tree decomposition in Fig.~\ref{fig:greedy:algo}(d).
   The hyperedge $\{C_4,C_6\}$ is still not contained in any bag in the resulting tree decomposition, which we fix by repeating the above process.
   In particular, we can add $\{C_4,C_6\}$ to the bag $\{X_5, X_6, C_6\}$.
   The resulting tree decomposition satisfies both properties (a) and (b) of Definition~\ref{defn:td}, hence the algorithm terminates.
   \qed
\end{ex}
\input{fig-greedy-algo}

\paragraph*{Exploiting symmetry}
The original color-coding paper~\cite{DBLP:journals/jacm/AlonYZ95} made an important observation that helps reduce the query complexity of the $k$-path query 
from $2^{O(k\log k)}$ down to $2^{O(k)}$. This is an exponential reduction in query complexity and helped answer an open question at the time: 
the $k$-path query, or more generally the bounded-treewidth subgraph isomorphism queries, 
can be solved in polynomial time for $k$ up to $O(\log n)$. 
In the dynamic programming algorithm that is used to evaluate the query (for example $\InsideOut$~\cite{faq} or Yannakakis~\cite{Yannakakis:VLDB:81}), the idea is 
to keep for each vertex only the (unordered) {\em sets} of colors it has
seen instead of the (ordered) {\em tuples} of colors it has encountered.

We can generalize this idea to our context as follows.
Consider the subproblem when $\InsideOut$ is about to eliminate a variable $Z$,
which is either a color variable $C_j$ or an input variable $X_i$. The
subproblem computes an intermediate result $R$ whose support is the set
$J^\pi_Z-\{Z\}$. 
To simplify notation, let $I$ denote the set of input variables in
$J^\pi_Z-\{Z\}$ and $K$ denote the set of color variables in $J^\pi_Z-\{Z\}$.
A tuple $\bm t_{J^\pi_Z-\{Z\}}$ 
in the intermediate result contains two sub-tuples: 
$\bm t_{J^\pi_Z-\{Z\}} = (\bm x_I, \bm c_K).$
Due to symmetry of the colorings, for a given $\bm x_I$ there may be several
tuples $\bm c_K$ for which $(\bm x_I, \bm c_K) \in R$ but many of the $\bm c_K$
are redundant in the following sense. Let $\bm c_K$ and $\bm c'_K$ be two
different color assignments to variables in $K$. Let $L$ be the set of color
variables other than those in $K$. 
Assume that the variables in $L$ have not been eliminated yet. 
Then, $\bm c_K$ and $\bm c'_K$ are said to be {\em equivalent} if, for any color
assignment $\bm c_L$ to variables in $L$, the assignment $(\bm c_L, \bm c_K)$ is
a proper coloring of the induced subgraph $\calG[L\cup K]$ if and only if
$(\bm c_L, \bm c'_k)$ is a proper coloring of the same induced subgraph.
The (slight) generalization of the idea from~\cite{DBLP:journals/jacm/AlonYZ95}
is that, if $(\bm x_I, \bm c_K)$ and $(\bm x_I, \bm c'_K)$ are both in the
intermediate result $R$, and if $\bm c_K$ and $\bm c'_k$ are equivalent, then we
only need to keep one of $\bm c_K$ or $\bm c'_K$, so only one tuple per equivalence class. This equivalence is akin to $\calH$-equivalence in~\cite{Koutris:TCS:17}.

It is straightforward to see that the above generalizes the idea
from~\cite{DBLP:journals/jacm/AlonYZ95}. The remaining question is how do we check
efficiently whether $\bm c_K$ and $\bm c'_K$ are equivalent? 
We next present a sufficient condition which can be verified efficiently.
Let $C_i \in L$ be an un-eliminated color variable not in $K$. A color $x \in
[c]$ is said to be a {\em forbidden color} for $C_i$ w.r.t the coloring $\bm c_K$ 
if the following holds. There is
a hyperedge $S \in \calA$ such that $C_i \in S$ and $S-\{C_i\} \subseteq K$.
Furthermore, the coloring $\bm c_K$ assigns the same color $x$ to all variables
in $S - \{C_i\}$. 
The set of all forbidden colors is called the {\em forbidden spectrum} of $C_i$
w.r.t. the coloring $\bm c_K$.
The following proposition is straightforward to verify:

\bprop
Two color tuples $\bm c_K$ and $\bm c'_K$ are equivalent if every color variable
$C_i \in L$ has precisely the same forbidden spectrum w.r.t. $\bm c_K$ and
w.r.t. $\bm c'_K$. 
\eprop

The ``identical forbidden spectrum'' condition can be verified in a brute-force
manner, every time we are about to insert a new tuple $(\bm x_I, \bm c_K)$ into
the intermediate relation $R$. When specialized to detecting a
$k$-path in a graph, this algorithm retains the $O(2^{O(k)})$-query complexity factor of the
original color-coding technique.

%% file: fig-H-H.tex
\begin{figure}[th!]
\newcommand{\drawvertexi}[4]
{
   \node[draw, circle, inner sep =.15] at (#1,#2) (#3) {#4};
}
\newcommand{\drawedgei}[5]
{
   \draw[#3](#1)--(#2) node[#4]{#5};
}
\begin{minipage}[b]{.5\linewidth}
{
\begin{tikzpicture}[scale=.6]
\foreach \i in {1,...,6}
   \drawvertexi{2*\i-2}{0}{X\i}{$X_{\i}$};
\drawedgei{X1}{X2}{blue, very thick}{midway,below}{$R_1$};
\drawedgei{X2}{X3}{blue, very thick}{midway,below}{$R_2$};
\drawedgei{X3}{X4}{blue, very thick}{midway,below}{$R_3$};
\drawedgei{X4}{X5}{blue, very thick}{midway,below}{$R_4$};
\drawedgei{X5}{X6}{blue, very thick}{midway,below}{$R_5$};
\end{tikzpicture}
}
\subcaption{The graph $\calH$}
\end{minipage}%
\begin{minipage}[b]{.5\linewidth}
{
\begin{tikzpicture}[scale=.6]
\foreach \i in {1,...,6}
   \drawvertexi{2*\i-2}{0}{X\i}{$X_{\i}$};
\foreach \i in {1,2,4,6}
{
   \drawvertexi{2*\i-2}{1.25}{C\i}{$C_{\i}$};
   \drawedgei{X\i}{C\i}{green!50!black, very thick}{midway,left}{};
}
\drawedgei{X1}{X2}{blue, very thick}{midway,below}{$R_1$};
\drawedgei{X2}{X3}{blue, very thick}{midway,below}{$R_2$};
\drawedgei{X3}{X4}{blue, very thick}{midway,below}{$R_3$};
\drawedgei{X4}{X5}{blue, very thick}{midway,below}{$R_4$};
\drawedgei{X5}{X6}{blue, very thick}{midway,below}{$R_5$};

\draw[red, very thick] (C1) to[out=30, in=150] (C4);
\draw[red, very thick] (C2) to[out=20, in=160] (C4);
\draw[red, very thick] (C4) to[out=20, in=160] (C6);
\end{tikzpicture}
}
\subcaption{The graph $\calH'$}
\end{minipage}%
\caption{The graphs $\calH$ and $\calH'$ from Example~\ref{ex:H:H'}.}
\label{fig:H:H'}
\end{figure}

%% file: fig-greedy-algo.tex
\tikzstyle{TDnode} = [draw, thick, ellipse, align=center, inner sep =.05cm]
\tikzstyle{TDedge} = [thick]
\colorlet{clr_Ci}{green!40!black}
\begin{figure}[th!]
\begin{minipage}[b]{\linewidth}
{
   \begin{tikzpicture}[scale=.9]
   \node[TDnode] at (0, 0) (bag1) {$X_1$\\$X_2$};
   \node[TDnode] at (2, 0) (bag2) {$X_2$\\$X_3$};
   \node[TDnode] at (4, 0) (bag3) {$X_3$\\$X_4$};
   \node[TDnode] at (6, 0) (bag4) {$X_4$\\$X_5$};
   \node[TDnode] at (8, 0) (bag5) {$X_5$\\$X_6$};
   \draw[TDedge] (bag1)--(bag2);
   \draw[TDedge] (bag2)--(bag3);
   \draw[TDedge] (bag3)--(bag4);
   \draw[TDedge] (bag4)--(bag5);
   \end{tikzpicture}
}
\subcaption{Tree decomposition of $\calH$}
\end{minipage}
\begin{minipage}[b]{\linewidth}
{
   \begin{tikzpicture}[scale=.9]
   \node[TDnode] at (0, 0) (bag1) {$X_1$\\$X_2$\\$\color{clr_Ci}C_1$\\$\color{clr_Ci}C_2$};
   \node[TDnode] at (2, 0) (bag2) {$X_2$\\$X_3$\\$\color{clr_Ci}C_2$};
   \node[TDnode] at (4, 0) (bag3) {$X_3$\\$X_4$\\$\color{clr_Ci}C_4$};
   \node[TDnode] at (6, 0) (bag4) {$X_4$\\$X_5$\\$\color{clr_Ci}C_4$};
   \node[TDnode] at (8, 0) (bag5) {$X_5$\\$X_6$\\$\color{clr_Ci}C_6$};
   \draw[TDedge] (bag1)--(bag2);
   \draw[TDedge] (bag2)--(bag3);
   \draw[TDedge] (bag3)--(bag4);
   \draw[TDedge] (bag4)--(bag5);
   \end{tikzpicture}
}
\subcaption{Adding {\color{clr_Ci}$C_i$} for every $X_i$ in the same bag}
\end{minipage}
\begin{minipage}[b]{\linewidth}
{
   \begin{tikzpicture}[scale=.9]
   \node[TDnode,draw=red] at (0, 0) (bag1) {$X_1$\\$X_2$\\$\color{clr_Ci}C_1$\\$\color{clr_Ci}C_2$\\$\color{clr_Ci}C_4$};
   \node[TDnode] at (2, 0) (bag2) {$X_2$\\$X_3$\\$\color{clr_Ci}C_2$};
   \node[TDnode,draw=blue] at (4, 0) (bag3) {$X_3$\\$X_4$\\$\color{clr_Ci}C_4$};
   \node[TDnode,draw=blue] at (6, 0) (bag4) {$X_4$\\$X_5$\\$\color{clr_Ci}C_4$};
   \node[TDnode] at (8, 0) (bag5) {$X_5$\\$X_6$\\$\color{clr_Ci}C_6$};
   \draw[TDedge] (bag1)--(bag2);
   \draw[TDedge] (bag2)--(bag3);
   \draw[TDedge,blue] (bag3)--(bag4);
   \draw[TDedge] (bag4)--(bag5);
   \end{tikzpicture}
}
\subcaption{Adding hyperedge $\{C_1, C_4\}$ to the bag $\{X_1,X_2,C_1,C_2\}$.
Note that now the bags containing $C_4$ form two disconnected subtrees: the {\color{red}red} and the {\color{blue}blue}.}
\end{minipage}
\begin{minipage}[b]{\linewidth}
{
   \begin{tikzpicture}[scale=.9]
   \node[TDnode,draw=red] at (0, 0) (bag1) {$X_1$\\$X_2$\\$\color{clr_Ci}C_1$\\$\color{clr_Ci}C_2$\\$\color{clr_Ci}C_4$};
   \node[TDnode] at (2, 0) (bag2) {$X_2$\\$X_3$\\$\color{clr_Ci}C_2$\\$\color{clr_Ci}C_4$};
   \node[TDnode,draw=blue] at (4, 0) (bag3) {$X_3$\\$X_4$\\$\color{clr_Ci}C_4$};
   \node[TDnode,draw=blue] at (6, 0) (bag4) {$X_4$\\$X_5$\\$\color{clr_Ci}C_4$};
   \node[TDnode] at (8, 0) (bag5) {$X_5$\\$X_6$\\$\color{clr_Ci}C_6$};
   \draw[TDedge] (bag1)--(bag2);
   \draw[TDedge] (bag2)--(bag3);
   \draw[TDedge,blue] (bag3)--(bag4);
   \draw[TDedge] (bag4)--(bag5);
   \end{tikzpicture}
}
\subcaption{Adding $C_4$ to bag $\{X_2,X_3,C_2\}$ to connect the two subtrees.}
\end{minipage}
\caption{Greedy algorithm for {\sf color amendment} applied to query from Example~\ref{ex:H:H'}.}
\label{fig:greedy:algo}
\end{figure}
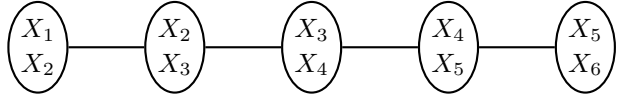
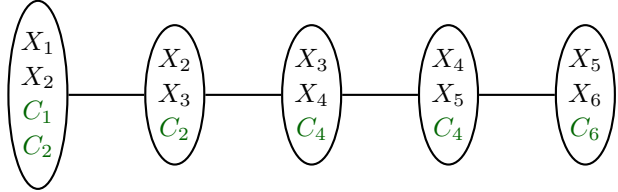
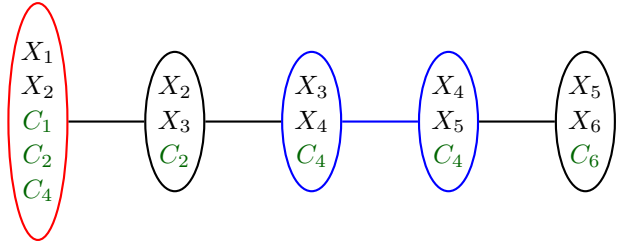
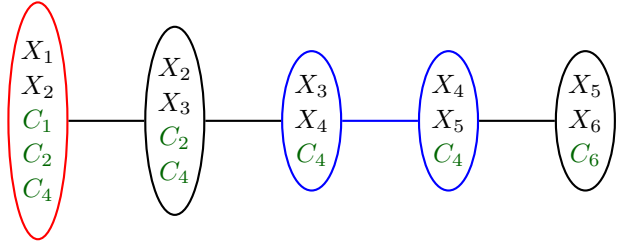